\documentclass[twocolumn,prb,showpacs,preprintnumbers,amsmath,amssymb]{revtex4}

\usepackage{graphicx}
\usepackage{dcolumn}
\usepackage{bm}
\usepackage{mdwlist}
\usepackage{textcomp}
\usepackage{hyperref}

\newcommand{\op}[1]{\hat #1}                                    
\newcommand{\mv}[1]{\langle#1\rangle}                           
\newcommand{\Mv}[1]{\left\langle#1\right\rangle}                
\newcommand{\Abs}[1]{\left| #1\right|}                          
\newcommand{\abs}[1]{| #1|}                                     
\newcommand{\R}{R_{\rm eo}}                                     

\begin{document}

\title{Main phase transition in lipid bilayers: phase coexistence and
line tension in a soft, solvent-free, coarse-grained model}

\author{Martin H{\"o}mberg}
\author{Marcus M{\"u}ller} \email{mmueller@theorie.physik.uni-goettingen.de}
\affiliation{Institut f{\"u}r Theoretische Physik, Georg-August-Universit{\"a}t,
Friedrich-Hund-Platz 1, 37077 G{\"o}ttingen, Germany}

\begin{abstract}
We devise a soft, solvent-free, coarse-grained model for lipid bilayer membranes. The non-bonded interactions take the form of a weighted-density functional which allows us to describe the thermodynamics of self-assembly and packing effects of the coarse-grained beads in terms of a density expansion of the equation of state and the weighting functions that regularize the microscopic bead densities, respectively. Identifying the length and energy scales via the bilayer thickness and  the thermal energy scale, $k_BT$, the model qualitatively reproduces key characteristics (e.g., bending rigidity, area per lipid molecules, and compressibility) of lipid membranes. We employ this model to study the main phase transition between the liquid and the gel phase of the bilayer membrane. We accurately locate the phase coexistence using free energy calculations and also obtain estimates for the bare and the thermodynamic line tension.
\end{abstract}

\maketitle
\section{Introduction}
\label{introduction}

Lipid bilayers are one of nature's most ingenious inventions.  \cite{Lodish03,Mouritsen05} They serve as a compartment to all cells, which form the building blocks of life, and they mediate the transport of molecules from the inside to the outside of cells. Many important properties of bilayer membranes involve collective phenomena, where a large number of interacting lipid molecules participate. Examples include the self-assembly of amphiphilic molecules into bilayer membranes, phase transitions between different phases or changes of the membrane topology, e.g., pore formation or fusion.\cite{Muller03b,Muller06} Computer simulations contribute to the understanding how these collective phenomena depend on the properties of the individual, constituent molecules.

Often, collective phenomena involve mesoscopic time- and length scales -- microseconds and \textmu m -- which are difficult to observe directly in experiments and which are at present beyond the scales that can be addressed by models with atomistic resolution. Therefore, several computational models have been developed where local, atomistic degrees of freedom have been integrated out.\cite{Brannigan06b,Venturoli06,Muller06} The reduced number of degrees of freedom of coarse-grained models and the softer interactions between the effective interaction centers opens up the opportunity to computationally address the mesoscopic scales involved in collective phenomena in complex biological matter. In order to design a coarse-grained model, first, one has to decide, which are the relevant degrees of freedom for the phenomena under study and which are to be integrated out. Second, one has to construct the effective interactions between the remaining degrees of freedom.  This construction is either performed systematically by explicitly tracing out the microscopic degrees of the freedom, or one invokes the concept of universality and uses a minimal set of interactions that is comprised only of those interactions, which are necessary to bring about the phenomena under study. The strength of those relevant interactions can be parameterized by comparing properties of the model to experimental data.

In this study we rely on the concept of universality. Instead of trying to reproduce chemical details of a specific lipid molecule, like it is done in atomistic or systematically coarse-grained simulations, we present a coarse-grained, solvent-free model for amphiphilic bilayers. Our model has some similarities with models used in self-consistent field calculations and it interpolates smoothly between lipid bilayers and polymeric membranes. Within the mean-field approximation of our model, there is a clear separation between the thermodynamic properties and the local fluid structure (i.e., packing effects) of the hydrophobic core of the bilayer membrane. We investigate the main phase transition between a liquid and a gel state of a self-assembled, one-component bilayer membrane and the line tension between domains. The main phase transition has been characterized in experiments for many lipids,\cite{Koynova98,Nagle00} and it has also been considered in many coarse-grained models of lipid bilayers.\cite{Ipsen90,Marrink04b,Marrink05,Kranenburg05,Lenz05,Cooke05,STEVENS04,Revalee08} The relation between the microscopic properties of the lipid molecules (e.g., the stiffness of the hydrocarbon tails and the fluid-like packing effects) and the macroscopic phase behavior, however, is only incompletely understood. Moreover, there are only very few attempts to accurately locate the phase boundaries.\cite{Mouritsen83} Problems arise from the hysteresis effects and metastability at the first-order transition, which seriously hamper the accurate determination of the location of the main phase transition by computer simulation. Additionally, the line tension between the fluid and gel phases has not been measured in coarse-grained models, which retain the notion of lipid molecules. This free energy of the domain boundaries between laterally coexisting phases has attracted abiding experimental interest.\cite{Allain04,Joannis06,Baumgart03,Tian07,Esposito07}

In order to describe the main phase transition, our coarse-grained model has to incorporate both (i) the minimal interactions that bring about the self-assembly into a bilayer membrane and, additionally, (ii) further details of the local inter- and intramolecular structure that give rise to the transition from a liquid to a gel phase. Thus, the following relevant properties are retained in our coarse-grained representation: (i) Each lipid comprises two different constituents, a hydrophobic tail and a slightly smaller hydrophilic head, which repel each other. These interactions drive the self-assembly into bilayer membranes. Since the lipid bilayer is typically surrounded by a solvent, the hydrophilic heads turn towards the solvent and the hydrophobic tails lump together in the bilayer's center. (ii) The hydrocarbon tails of the lipids are characterized by a finite length, a limited conformational flexibility and a finite excluded volume diameter.  These interactions give rise to a crystalline packing of the molecules in the gel phase. Incorporating both aspects, our coarse-grained model bridges between minimal coarse-grained representations, which only capture the universal aspects of self-assembly, and systematically coarse-grained models that have been explicitly derived from an atomistic model.

Our manuscript is arranged as follows: In Sec.~\ref{model} we describe our solvent-free, coarse-grained model with soft interactions and provide details of the Multibody Dissipative Particle Dynamics (MDPD) simulations. A technical description of the symplectic integration algorithm for simulating in an ensemble with constant tension is deferred to Appendix~\ref{integrator}. The subsequent section, Sec.~\ref{properties}, demonstrates the self-assembly of lipids into bilayer membranes. Several static and dynamic properties of our model, such as the bending rigidity, the bilayer density profile, and the molecular diffusion coefficient are measured.  In Sec.~\ref{main-phase-transition} the main phase transition is studied. We use Umbrella Sampling (US)\cite{umbrella,Ferrenberg89,Virnau04b} to restrain the fluctuations of an orientational order parameter. Changing the order parameter, we reversibly transform the fluid into a gel phase and obtain the concomitant free-energy profile by the Weighted Histogram Analysis Method (WHAM).\cite{Ferrenberg88,Ferrenberg89,WHAM,Souaille01,Chipot07} The bilayer configurations along the reversible path are discussed.  Sec.~\ref{line-tension} describes the measurement of the thermodynamic line tension between gel and fluid domains, extracted from the free-energy profile, and the bare line tension, computed from the fluctuation spectra of the domain boundaries. The relation between these two properties is discussed in Appendix \ref{capillary-wave-appendix}. The paper concludes with a summary and an outlook in Sec.~\ref{conclusions}. 


\section{Model and technique}
\label{model}

\subsection{Model}
We consider a coarse-grained model for the simulation of lipid bilayer membranes. Our system contains $n$ lipid molecules that are represented by linear bead-spring chains comprising $N=16$ effective interaction centers, which are either hydrophobic (``$A$'') or hydrophilic (``$B$''). The ratio of $A$-beads, $N_A$, in a lipid is defined by the asymmetry parameter $f$, so that $N_A = fN$ and $N_B = (1-f)N$, respectively. The beads are connected by harmonic springs with spring constant $k_s$. Additionally, we apply a bond-angle potential between every three successive beads $i-1,i,i+1$ with constant $k_b$ to stiffen the lipids. Thus, the intramolecular, bonded interactions of a single lipid are given by
\begin{equation}
\label{u-b}
\frac{\mathcal{H}_{\text{b}}}{k_BT} = \sum\limits_{i=1}^{N-1} \frac{k_s}{2} \left[\mathbf{r}_{i+1}-\mathbf{r}_i\right]^2 + \sum\limits_{i=2}^{N-1} k_b \left[1 - \cos\theta_i\right],
\end{equation}
where $\theta_i$ is the angle between the vectors $\left(\mathbf{r}_i - \mathbf{r}_{i-1}\right)$ and $\left(\mathbf{r}_{i+1}-\mathbf{r}_i\right)$. The thermal energy, $k_BT$, serves as the unit of energy in our model. Although lipid molecules are characterized by several length scales, we use the root-mean-squared end-to-end distance, $\R = \mv{\left(\mathbf{r}_1-\mathbf{r}_N\right)^2}^{1/2}$, of lipids that are only subjected to the bonded interactions, ${\cal H}_{\rm b},$\cite{KADV,Daoulas06b} as the characteristic dimension of the bilayer. It can be pictured as the head-to-tail length of a single lipid in vacuum. The use of $\R$ to specify the molecular extension is rooted in polymeric membranes, where the polymer conformations are characterized by this single length scale.\cite{KADV} Its value, in turn, depends on the values of $N$, $k_s$, and $k_b$. The bond stiffness restricts the conformational fluctuations of the amphiphilic molecules such that the average molecular size and its shape fluctuations are controlled by the parameters of the model. The actual size of a lipid molecule, of course, is influenced by the interactions with its neighbors, e.g., it differs in the liquid and the gel phase.

Since on large length scales a bilayer membrane can be conceived as a thin, two-dimensional sheet embedded in a three-dimensional volume, most of the volume is occupied by solvent. Although the solvent acts as a transport medium in a plethora of biological processes and mediates the self-assembly, drastically simplifying its representation or even integrating out the solvent altogether offers a potentially huge reduction in the number of the degrees of freedom.\cite{Lenz05,Gao07,Drouffe91,NOGUCHI01_2,FARAGO03,Brannigan04,Cooke05,Revalee08} By integrating out the degrees of freedom of the solvent, the original interactions of the underlying model containing the explicit solvent molecules are turned into effective interactions. These depend on the thermodynamic state, at which the elimination of the explicit solvent has been performed. Thus, the non-bonded interactions are free energies and care has to be exerted when extracting thermodynamic properties.\cite{Louis02}

In the following we employ a solvent-free model to study thermodynamic equilibrium properties. Thus, hydrodynamic interactions, that are mediated by the solvent, are irrelevant. The non-bonded interactions are accounted for by a phenomenological Ansatz for the excess free energy. Specifically, we use an expansion up to third order for the non-bonded excess free energy in terms of the dimensionless, weighted densities of the molecules.\cite{KADV}
\begin{equation}
\label{u-nb1}
\frac{\mathcal{H}'_{\text{nb}}}{k_BT} = \int \frac{{\rm d}^3r}{\R^3} \, \rho_\alpha \left[ \frac{ v_{\alpha\beta}}{2}\rho_\beta+ \frac{w_{\alpha\beta\gamma}}{3} \rho_\beta\rho_\gamma \right]
\end{equation}
A summation over all Greek indices that occur twice is implied and the integration extends over the whole volume of the simulation box. The term in the bracket denotes the excess free energy per particle. The weighted densities are related to the explicit particle coordinates via a weighted average over a small volume. The details of this procedure are discussed below. Here we only note that, once the weighted densities are specified in terms of the microscopic particle coordinates, the Hamiltonian \eqref{u-nb1} becomes a function of the explicit particle coordinates and the properties of the coarse-grained model can be studied by computer simulation.

Within the mean-field approximation, the properties of the particle-based simulation model coincide with the results of a density functional theory (DFT) calculation using the excess free energy functional $\mathcal{H}'_{\text{nb}}[\rho_A,\rho_B]$. In particular, within the mean-field approximation, thermodynamic and structural properties decouple.\cite{Muller00i} The thermodynamic properties of a spatially homogeneous system, e.g., the equation of state, are dictated by the seven expansion coefficients, $v_{\alpha\beta}$ and $w_{\alpha\beta\gamma}$. The local structure of the liquid, in turn, is encoded in the definition of the weighted densities.

The advantages of these DFT-based, non-bonded interactions are twofold: On the one hand, Eq.~\eqref{u-nb1} can be generalized in a systematic way to accommodate more sophisticated equations of state. In the present work, we use a third-order expansion \cite{Muller02b,KADV} because this is the simplest form capable of describing all six, qualitatively different types of phase diagrams that a compressible binary system exhibits according to the classification of \citeauthor{vankonynenburg80},\cite{vankonynenburg80} i.e., it suffices to capture all qualitative features of the interplay between liquid-vapor phase separation and demixing of two species. Moreover, by virtue of its simplicity, the second- and third-order coefficients are straightforwardly related to the density and compressibility of a homogeneous liquid and the incompatibility between hydrophobic and hydrophilic entities. This relation imparts a transparent physical interpretation onto the coefficients. The density expansion also allows for a systematic generalization to systems comprised of more than two different species.\cite{Wang09a,Wang09b} This situation naturally arises in the study of more complex systems. On the other hand, the weighted densities encode local structural information. Altering the definition of the weighted density, we are able to describe lipid bilayer membranes, which exhibit pronounced packing effects on the length scale of an effective interaction center, or polymersomes that are comprised of long, flexible, amphiphilic polymers and, typically, do not form gel phases.

We discuss how to choose the expansion coefficients and the definition of the weighted densities in turn.

\subsubsection{Thermodynamic coefficients of the third-order density expansion}
Formally, we consider the system of amphiphiles and solvent on the mesoscopic scale of a coarse-grained interaction center as an incompressible, dense liquid with bulk density $\rho_{\rm o}$. Knowing the local densities of amphiphiles, one can reconstruct the solvent density by assuming that the total system of solvent and amphiphiles is nearly incompressible and integrate out the degrees of freedom associated with the solvent.\cite{Louis02,Muller06} This gives rise to effective interactions and the incompressibility constraint generates multi-body interactions. The occurrence of multi-body interactions is natural in the course of coarse-graining and it would also arise during a systematic coarse-graining procedure where microscopic degrees of freedom are explicitly integrated out.

The coefficients $v_{AA}$ and $w_{AAA}$ dictate the properties of the hydrophobic species in contact with the solvent. In a solvent-free model, the hydrophobic species forms a dense liquid that coexists with a vapor phase, which represents the solvent. Since the solubility of amphiphiles in the solvent is vanishingly small, the (osmotic) pressure of the vapor phase, which coexists with the liquid, vanishes, $P \approx 0$. Using the mean-field equation of state for the pure $A$-component
\begin{equation}
\frac{P \R^3}{k_BT} \approx \rho_A + \frac{v_{AA}}{2} \rho_A^2 + \frac{2w_{AAA}}{3} \rho_A^3
\label{eos}
\end{equation}
we obtain for the molecular density, $\rho_{\rm coex}$, of the liquid with $P=0$
\begin{equation}
\rho_{\rm coex} \approx - \frac{3v_{AA}}{4w_{AAA}}
\label{r-coex}
\end{equation}
and for the dimensionless, inverse compressibility
\begin{eqnarray}
\kappa N &\equiv &\frac{\R^3}{ \kappa_T\rho_{\rm coex} k_BT} = v_{AA} \rho_{\rm coex} + 2 w_{AAA} \rho_{\rm coex} \\
\mbox{with} && \kappa_T \equiv - \frac{1}{V} \left. \frac{\partial V}{\partial P}\right|_T
\label{kN}
\end{eqnarray}
respectively. In both cases we have neglected the contribution of the first term in the equation of state (\ref{eos}) that corresponds to an ideal gas. These approximate expressions provide a simple physical interpretation of the expansion coefficients. We will present our results as a function of $\kappa N$ and $\rho_{\rm coex}$ using the dependencies
\begin{equation}
v_{AA} = -2 \frac{\kappa N +3}{\rho_{\text{coex}}}
\qquad\text{and}\qquad
w_{AAA} = \frac{3}{2} \frac{\kappa N + 2}{\rho^2_{\text{coex}}}.
\label{tune}
\end{equation}
We use $\rho_{\rm coex}$ as control parameter to study the main phase transition between a fluid and a gel phase. At large $\rho_{\rm coex}$ molecules strongly overlap, packing effects are small, and the system is in the fluid phase. This behavior is typical for polymersomes, where a coarse-grained bead is comprised of many atomistic units or for high temperatures, where the soft, non-bonded interactions are weak compared to the thermal energy scale. A decrease of $\rho_{\rm coex}$, in turn, corresponds to an increase of the repulsive, third-order interactions (cf.~Eq.~(\ref{tune})), which gives rise to a transition from the fluid to the gel phase.

The coefficient, $v_{AB}$, sets the strength of the interactions between $A$ and $B$ beads. It is related to the Flory-Huggins parameter, $\chi N$, via
\begin{equation}
v_{AB} =  \frac{\chi N}{\rho_{\text{coex}}} + \frac{1}{2}\left(v_{AA} + v_{BB}\right).
\end{equation}
The dimensionless, invariant quantity, $\chi N$, measures the incompatibility between hydrophilic and hydrophobic species. $v_{BB}$ and $w_{BBB}$ are chosen, such that the hydrophilic beads are in a good solvent, i.e., their interactions are purely repulsive, $v_{BB}=0.1$ and $w_{BBB}=0$. The mixed, third-order coefficients, $w_{AAB}$ and $w_{ABB}$, do not influence the qualitative behavior and, for simplicity, we set $w_{AAA}=w_{AAB}=w_{ABB}$.

Four phenomenological parameters describe the thermodynamics of our soft, solvent-free, coarse-grained model: $\rho_{\rm coex}$, $\kappa N, \chi N$ and $\R$, which parameterize (i) the density and (ii) the limited compressibility of the hydrophobic interior, (iii) the incompatibility between hydrophilic and hydrophobic beads, and (iv) the spatial extension of a lipid molecule. All these parameters are directly related to experimentally accessible quantities and our model can be related to a specific system by matching these four parameters of our coarse-grained model to experimental data.

For instance, we estimate the order of magnitude of $\kappa N$ from the bulk properties of an  alkane liquid. Using the isothermal compressibility under standard conditions $\kappa_T=0.955~\text{GPa}^{-1}$ for $n$-Dodecane,\cite{Khasanshin03} its bulk mass density $\rho_m=748.8~\text{kg/m}^3$, and its molar mass $m=170.34~\text{g/mol}$, we obtain $\kappa N = m/\left(\kappa_T \rho_m k_BT\right) \approx 98$.

\subsubsection{Weighted densities}
For lipid bilayer membranes we seek for weighted densities that yield a phase diagram with the biologically important fluid phase and, additionally, various gel phases. Analytical studies have suggested that the phase behavior of lipid bilayers is dominated by packing effects due to the excluded volume of the hydrophobic tails.\cite{Mouritsen91} In our model, we can draw on the vast knowledge of liquid-state theory to control the degree of packing effects and local structure of the fluid in order to tailor the weighted densities such that the fluid exhibits pronounced packing effects. 

The dimensionless, microscopic densities, $\hat \rho_\alpha(\mathbf r)$,  of hydrophilic and hydrophobic species are functions of the explicit coordinates of the effective interaction centers

\begin{equation}
\label{bead-density}
\hat \rho_\alpha(\mathbf r) = \frac{\R^3}{N} \sum\limits_{i=1}^{nN} \delta(\mathbf r_i - \mathbf r) \delta_{\alpha t(i)},
\end{equation}
where $t(i) \in \{A,B\}$ denotes the species of bead $i$. The prefactor has been chosen such that the molecular density does not depend on the number of interactions centers per molecule, $N$. In order to regularize the $\delta$-function in the excess free-energy functional of non-bonded interactions, Eq.~\eqref{u-nb1}, we use a weighted-density approximation\cite{WDA,Vanswol91,Yethiraj98,Muller03d} and define coarse-grained densities 

\begin{figure}[tb]
\includegraphics[clip,width=\linewidth]{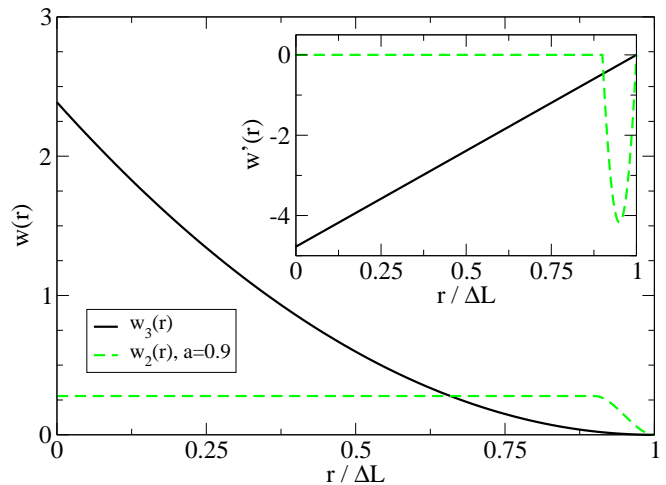}
\caption{\label{fig:wf}
The two weighting functions from eqs.~\eqref{w-cs} and \eqref{w-pb} with $a=0.9 \Delta L$. The inset shows the derivatives of these functions, which are proportional to the non-bonded forces between two beads.}
\end{figure}
\begin{equation}
\label{w-bead-density}
\bar\rho_{m\alpha}(\mathbf r) = \frac{\R^3}{N} \sum\limits_{i=1}^{nN} w_m\left(\Abs{\mathbf r_i - \mathbf r}\right) \delta_{\alpha t(i)}
\end{equation}
by convoluting the microscopic, molecular density, $\hat \rho_\alpha(\mathbf r)$ with weighting functions, $w_m$. We require that the weighting functions are differentiable, vanish for $r \ge \Delta L$, and are normalized, i.e., $\int {\rm d}^3r\,w_m(\abs{\mathbf{r}}) = 1$. Liquid-state theory for simple liquids \cite{WDA,Vanswol91} as well as integral equation theory \cite{Weeks98,Katsov01} indicate that it is important to use different weighting functions to represent the harsh, short-ranged repulsion in a liquid and the soft, longer-ranged attractions. The second-order terms in Eq.~\eqref{u-nb1} typically correspond to attractive interactions and the third-order terms to repulsions. Therefore, we use different weighting functions, $w_2$ and $w_3$, for the second and third-order contributions. Both weighting functions are plotted in Fig.~\ref{fig:wf}. The longer-ranged weighting function, $w_2$, consists of a constant part for $r \le a$ and a cubic spline for $a < r \le \Delta L$ with $0<a<\Delta L$, given by
\begin{equation}
\label{w-cs}
w_2(r)=\\
A \begin{cases}
\left(\Delta L-a\right)^3, & r \le a \\
2r^3-3(a+\Delta L)r^2+6a\Delta Lr \cdots&\\
\cdots -3a \Delta L^2+ \Delta L^3, & r < \Delta L
\end{cases},
\end{equation}
$A = -15/\left[2\pi(2a^6-3a^5\Delta L+3a \Delta L^5-2 \Delta L^6)\right]$ is a normalization constant. In the following we use $a = 0.9~\Delta L$. It is used for the mainly attractive, second-order terms. The weighting function for the repulsive interactions, $w_3$, is the standard choice in Dissipative Particle Dynamics models.\cite{DPD2b} 
\begin{equation}
\label{w-pb}
w_3(r) = \frac{15}{2\pi \Delta L^5} \left(\Delta L-r\right)^2.
\end{equation}
It only possesses positive Fourier modes. Negative Fourier modes of pair-wise, repulsive interactions give rise to cluster-crystallization in dense liquids of soft particles. \cite{LikosCC07,Mladek05,Mladek07} Our choice of weighting functions avoids the formation of cluster-crystals in the range of parameters investigated in the following.

Using Eqs.~\eqref{bead-density} and \eqref{w-bead-density}, we rewrite the non-bonded interactions in the form
\begin{equation}
\frac{\mathcal{H}_{\text{nb}}}{k_BT} = \int \frac{{\rm d}^3r}{\R^3}\,\hat\rho_\alpha(\mathbf r)
\left[ \frac{v_{\alpha\beta}}{2}\bar\rho_{2\beta}(\mathbf r)+
\frac{w_{\alpha\beta\gamma}}{3}\bar\rho_{3\beta}(\mathbf r)\bar\rho_{3\gamma}(\mathbf r)\right]
\label{u-nb}
\end{equation}
which takes the form of a weighted-density functional.\cite{Curtin85,Vanswol91,Yethiraj98,Muller03d} The density-functional form of this coarse-grained interaction free energy controls local correlations, e.g., packing effects. Their length scale is set by the spatial extent of the non-bonded interaction, $\Delta L$. Unlike density-functional theory, however, we obtain the properties not by minimizing the density functional but we use density-functional-inspired interactions in our soft, coarse-grained model whose properties are studied by computer simulation. In this way, long-range fluctuations, e.g., undulations of the bilayer membrane, are accounted for.

Finally, we note that weighted densities which give rise to strong packing effects deteriorate the quality of the mean-field approximation and, consequently, the decoupling between the thermodynamic properties (e.g., compressibility and coexistence density) and the liquid structure breaks down. Therefore, the model parameters, $\rho_{\rm coex}$ and $\kappa N$, are not identical to the density in the hydrophobic interior of the bilayer and its inverse compressibility. Nevertheless, the approximate equations, \eqref{r-coex} and \eqref{kN}, are a useful guide for constructing the model.

\subsection{Simulation technique}
We applied Multibody Dissipative Particle Dynamics (MDPD)\cite{Pagonabarraga01,Trofimov02,Warren03} to integrate the stochastic equations of motion. In MDPD the force $\mathbf F_i$ acting on each bead, $i$, consists of three terms,
\begin{equation}
\mathbf F_i = \sum\limits_{j \neq i}^{nN} \mathbf F^C(\mathbf r_{ij}) +
\mathbf F^D(\mathbf r_{ij}, \mathbf v_{ij}) + \mathbf F^R(\mathbf r_{ij}).
\end{equation}
Here $\mathbf r_{ij} = \mathbf r_i - \mathbf r_j$, and $\mathbf F^C(\mathbf r_{ij}) = \mathbf F_{\text{b}}^C(\mathbf r_{ij}) + \mathbf F_{\text{nb}}^C(\mathbf r_{ij})$ is the pair-wise, conservative force. The contributions from the bonded interactions, $\mathbf F_{\text{b}}^C(\mathbf r_{ij})$ are obtained by taking the derivative of the potential energy in Eq.~\eqref{u-b} with respect to the coordinates of the beads. 

The non-bonded forces, $\mathbf F_{\text{nb}}^C(\mathbf r_{ij})$, stem from the density-dependent Hamiltonian \eqref{u-nb}. We rewrite Eq.~\eqref{u-nb} in a computationally convenient form using the expressions for the microscopic and weighted densities.
\begin{equation}
\frac{\mathcal{H}_{\text{nb}}}{k_BT} 
= 
\sum\limits_i \delta_{\alpha t(i)}
\left[
\frac{v_{\alpha\beta}}{2N} \bar \rho_{2\beta}(\mathbf r_i) +
\frac{w_{\alpha\beta\gamma}}{3N} \bar\rho_{3\beta}(\mathbf r_i) \bar\rho_{3\gamma}(\mathbf r_i)
\right] 
\end{equation}
Taking the negative derivative of $\mathcal{H}_{\text{nb}}$ with respect to $\mathbf r_i$, we obtain
\begin{eqnarray}
\mathbf F_{\text{nb},i}^C &=& -\frac{\partial}{\partial \mathbf r_i} \mathcal{H}_{\text{nb}} \\
& = & \sum\limits_j \mathbf{\hat r}_{ji} \biggl[ v_{t(i)t(j)} w_2'\bigl(\Abs{\mathbf r_j - \mathbf r_i}\bigr) \label{f-nb} \\
&&   \hspace*{1cm} + \frac{2 w_{t(i)t(j)\alpha}}{3} w_3'\bigl(\Abs{\mathbf r_j - \mathbf r_i}\bigr)
\Bigl(\bar\rho_{3\alpha}(\mathbf r_i) + \bar\rho_{3\alpha}(\mathbf r_j) \Bigr)
\biggr] \nonumber
\end{eqnarray}
Thus, the total non-bonded force is decomposed into a sum of pair-wise forces, $\mathbf F_i^C = \sum_j \mathbf F_{ij}$.

The dissipative force, $\mathbf F^D(\mathbf r_{ij}, \mathbf v_{ij})$, and the random force, $\mathbf F^R(\mathbf r_{ij})$, are used to obtain a canonical ensemble, in which the temperature is constant. They have the same cutoff, $\Delta L$, i.e. they vanish for $r_{ij} = \Abs{\mathbf r_{ij}} \ge \Delta L$. For $r_{ij} < \Delta L$ they are given by the DPD form:\cite{DPD1,DPD2}
\begin{eqnarray}
\mathbf F^D(\mathbf r_{ij}, \mathbf v_{ij}) &=& -\gamma \omega^D(r_{ij})(\mathbf v_{ij} \cdot \hat{\mathbf r}_{ij})\hat{\mathbf r}_{ij} \\
\mathbf F^R(\mathbf r_{ij}) &=& \xi \omega^R(r_{ij}) \theta_{ij} \hat{\mathbf r}_{ij}
\end{eqnarray}
The friction constant, $\gamma$, is related to the noise coefficient, $\xi$, by the fluctuation dissipation theorem, $\xi^2=2 k_BT \gamma$. $\theta_{ij}$ is a stochastic variable with mean, $\mv{\theta_{ij}}=0$, and covariance, $\mv{\theta_{ij}(t)\theta_{kl}(t')}= (\delta_{ij}\delta_{kl}+\delta_{il}\delta_{jk})\delta(t-t')$.  The random numbers are drawn from a uniform distribution,\cite{Dunweg1991} and the standard weighting functions for DPD\cite{Warren1995}
\begin{equation}
\left[\omega^R(r)\right]^2 = \omega^D(r) =
\begin{cases}
1-r/\Delta L, &  r<\Delta L \\
0, & r \ge \Delta L
\end{cases}
\end{equation}
are employed. In the following we use $1~\R \equiv 3.5~\Delta L$.

Several different thermodynamic ensembles have been used in the course of our study. Some simulations have been performed in the canonical ensemble (NVT) using the standard velocity-Verlet integration scheme with a time step of $\Delta t=0.005~\tau$.\cite{Allen87} Most of the simulations have employed an ensemble where the area of the lipid bilayer fluctuated, such that the lateral pressure vanished, i.e. $P_t=0$. The height of the simulation box $L_x$ in the direction normal to the bilayer was kept at a fixed value. We refer to this thermodynamic ensemble as the ``$NP_tT$'' ensemble. Details of the symplectic integration algorithm for this extended ensemble\cite{Kolb99} are given in Appendix~\ref{integrator}. $P_t=0$ and $P \approx 0$ imply that the bilayer is in a state of vanishing mechanical tension, $\Sigma$. The simulations have been performed by a parallel DPD program employing the force-decomposition algorithm devised by Plimpton.\cite{PLIMPTON95}


\section{Self-assembly and bilayer properties}
\label{properties}
In this section we demonstrate that the lipid molecules self-assemble into various morphologies, and we compile several static and dynamic properties of the soft, coarse-grained model. 

\subsection{Self-assembly}
\begin{table*}
\begin{minipage}{\textwidth}
\caption{Morphologies observed during self-assembly from a disordered starting configuration}
\begin{ruledtabular}
\begin{tabular}[b]{lrrr|cccccc}
$\rho_{\text{coex}}$ & $\kappa N$ & $\chi N$ & $k_b$ &
$10-6$\footnotemark[1] & $11-5$ &
$12-4$ & $13-3$ & $14-2$ & $15-1$ \\ \hline
15 & 100 & 40 & 0 & c\footnotemark[2] & b & b & i & i & i \\
15 & 100 & 40 & 5 & s & b & b & b & i & i \\
15 & 100 & 60 & 0 & s & b & b & b & i & i \\
15 & 100 & 60 & 5 & s & b & b & i & i & i \\
18 &  80 & 20 & 5 & s & w & w \\
18 &  80 & 30 & 5 & s & c & w \\
18 &  80 & 40 & 5 & s & c & w \\
20 & 100 & 20 & 5 & s & c & b & b \\
20 & 100 & 30 & 5 & s & c & b & i \\
20 & 100 & 40 & 5 & s & s & c & b \\
20 & 100 & 50 & 5 & s & s & w & b \\
\end{tabular}
\end{ruledtabular}
\footnotetext[1]{this notation means $N_A=10, N_B=6$}
\footnotetext[2]{s: spherical micelles, c: cylindrical micelles, w: wormlike
micelles, b: bilayer, i: bilayer with hydrophilic inclusions}
\label{tab:self-assembly}
\end{minipage}
\end{table*}
To study the self-assembly as a function of the molecular stiffness, we have used three different sets of $k_s$ and $k_b$ (cf.~Eq.~\eqref{u-nb}): (i) $k_s=3.673$ and $k_b=0$ for flexible lipids without any bond-angle potential, (ii) $k_s=19.0$ and $k_b=5$ representing lipids with a moderate stiffness, and (iii) $k_s=29.4$ and $k_b=10$ parameterizing lipids with a high stiffness. We have explored various values of the coarse-grained parameters, $20 \leq \chi N \leq 100$, $50 \leq \kappa N \leq 500$, and $15 \leq \rho_{\text{coex}} \leq 40$, as well as different lengths of the hydrophobic tails $N_A$ and the hydrophilic heads $N_B$ with $N_A+N_B=N=16$. We have performed all simulations in the $NP_tT$-ensemble with $\Sigma=0$ and have used the same initial configuration comprised of $n=1600$ lipids in a box of lengths $L_x=50~\Delta L$, and $L_y(t=0)=L_z(t=0)=30~\Delta L$. The lipids were randomly distributed over the lower half of the box, i.e. $0 < x < L_x/2$ to avoid the formation of multiple bilayers.

\begin{figure}[tb]
\includegraphics[clip,width=\linewidth]{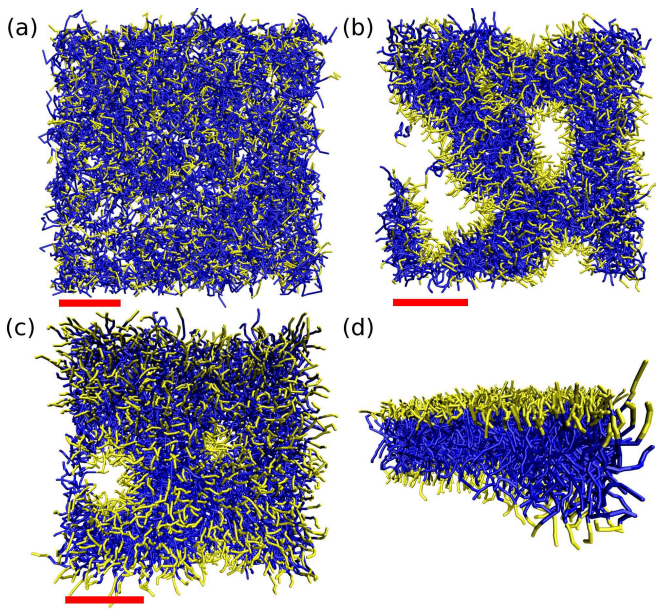}
\caption{\label{fig:self-assembly}
Self-assembly of the system $\rho_{\text{coex}}=17, \kappa N=100, \chi N=30$, $N_A=12$ hydrophobic (blue), and $N_B=4$ hydrophilic beads (yellow). From the initial configuration (a) broad wormlike micelles form (b). They coalesce forming a bilayer with several pores (c). These pores close slowly and a continuous bilayer is formed (d). It happened frequently that up to $5~\%$ of the lipids stayed at first in the gas phase, and formed after some time one micelle.  (a) $t=0$, (b) $40~\tau$, (c) $200~\tau$, (d) $800~\tau$. The red scale bar denotes $2~\R$. Created with VMD.\cite{VMD}}
\end{figure}
Depending on the parameter set, the lipids self-assemble within $\Delta t \le 500~\tau$ to one of the following morphologies: (i) spherical micelles, (ii) cylindrical micelles, (iii) wormlike micelles, (iv) bilayers, or (v) bilayers with hydrophilic inclusions. A typical pathway of the self-assembly of a bilayer is shown in Fig.~\ref{fig:self-assembly}. The results obtained with other parameter sets are compiled in Tab.~\ref{tab:self-assembly}. Bilayers form for $N_A \ge 11$, and inverted structures, i.e., bilayers with hydrophilic inclusions, form for $N_A \ge 13$. For $k_b=0,5,10$ we observe only the fluid phase, a fluid and a gel phase, and only a gel phase, respectively. For $\chi N \ge 50$ wormlike or cylindrical micelles predominantly form, whereas for $\chi N < 20$ the incompatibility between hydrophilic and hydrophobic beads becomes so small, that no clear separation between hydrophilic and hydrophobic regions is visible.

The observed sequence of morphologies is consistent with the geometrical arguments put forward by Israelachvili.\cite{Israelachvili91} For $N_A<11$ the amphiphiles have a conical shape, so that only micelles occur irrespective of the other parameters. $N_A=11$ and $N_A=12$ result in an almost cylindrical shape of the lipids, so that bilayers form. When the hydrophilic heads decrease in size, $N_A>12$, inverted morphologies appear. With increasing $\rho_{\text{coex}}$ each coarse-grained bead interacts with more neighbors, so that the mean-field approximation becomes more accurate and fluid-like packing effects weaker. This marks the crossover to polymeric membranes, where the chain number density is typically higher than in lipid bilayer membranes and only a fluid phase is stable.\cite{KADV}

The spatial extension of a lipid molecule is of the order $\R$, but the fluctuations around this mean value are largely influenced by $k_b$. The value, $k_b=0$, corresponds to fully flexible molecules and the shape fluctuations are of the same order of magnitude as the lipid's size, i.e., the conformations resemble a self-avoiding random walk. For $k_b=10$ the lipids are strongly elongated and they behave like rods. This gives rise to nematic, liquid crystalline structure of the self-assembled bilayers.

\subsection{Bilayer properties}
\begin{table*}
\begin{center}
\caption{Static and dynamic properties for $\kappa N=100$, $\chi N=30$, $n=4680$}
\begin{ruledtabular}
\begin{tabular}[b]{lllrrr}
& & & $\rho_{\text{coex}}=40$ & $\rho_{\text{coex}}=17$ & $\rho_{\text{coex}}=17$ \\
& & & Fluid ($L_\alpha$) & Fluid ($L_\alpha$) & Gel ($L_\beta$) \\ \hline
Area & $\mv{A}$ & $[\R^2]$ & $63.7(1)$ & $109.6(1)$ & $99.7(1)$ \\
Area Compressibility & $k_A$ & $[10^{-3}\R^2/k_BT]$ & $4.17(1)$ & $3.53(1)$ & $0.14(1)$ \\
Area per Lipid & $\mv{a}$ & $[10^{-2}\R^2]$ & $2.722$ & $4.684$ & $4.262$ \\
Bulk Density & $\rho_A$ & $[1/\R^{3}]$ & $42.9(1)$ & $22.4(1)$ & $23.6(1)$ \\
Width of Hydrophobic Layer & $w$ & $[\R]$ & $1.21(1)$ & $1.42(1)$ & $1.53(1)$ \\
Total Thickness & $t$ & $[\R]$ & $1.63(1)$ & $1.87(1)$ & $2.05(1)$ \\
Aspect Ratio & $w/\sqrt{\mv{a}}$ & $[1]$ & $7.3$ & $6.5$ & $7.4$\\
Bending Rigidity (spectrum) & $\kappa$ & $[k_BT]$ & $19(1)$ & $15(1)$ & -- \\
Bending Rigidity (from $k_A$) & $\kappa$ & $[k_BT]$ & $13$ & $21$ & -- \\
Molecular Diffusion Constant & $D$ & $[\R^2/\tau]$ & $1.5(1)\cdot 10^{-3}$ & $5.0(1)\cdot 10^{-4}$ & $2.6(1)\cdot 10^{-7}$ \\
\end{tabular}
\end{ruledtabular}
\label{tab:prop}
\end{center}
\end{table*}
In the following, we focus on the parameter set, $N_A=12$, $N_B=4$, $k_s=19.0$, and $k_b=5$, which gives rise to the spontaneous formation of bilayer membranes. The non-bonded interactions are set to $\kappa N=100, \chi N=30$ and $\rho_{\text{coex}}=17$ or $40$. The longest simulation runs lasted $\Delta t \approx 10^4~\tau$. We used pre-assembled bilayers as initial configurations with $n=4680$ lipids. In the case of $\rho_{\text{coex}}=17$ two different initial configurations have been used -- one in the liquid phase, $L_\alpha$, and one in the gel phase, $L_\beta$. The former one has also been employed as initial configuration for $\rho_{\text{coex}}=40$. We compiled the obtained properties in Tab.~\ref{tab:prop}.

\subsubsection{Density profiles}
Stable fluid membranes in a solvent form a bilayer structure. The hydrophilic head groups on the outside favor contact with the solvent, and the tails constitute the bilayer's hydrophobic interior, which is shielded from the solvent. This lamellar structure becomes visible in the molecular density profile, which has been recorded separately for the two leaflets. To avoid a broadening of these profiles by thermal undulations, the bilayer has been subdivided laterally into small cells of size, $\Delta L \times \Delta L$. In each cell, the local bilayer position has been determined and the profiles have been averaged with respect to this local bilayer position over all cells and along the trajectory. As noted above, packing effects result in a deviation of the density of $A$-beads in the hydrophobic core from the estimate of the coexistence density, $\rho_{\rm coex}$, provided by mean-field theory. The width, $w$, of the hydrophobic core is estimated in both phases by measuring the full width at half maximum (FWHM). The total bilayer thickness, $t$, is the distance from the center of mass of the hydrophilic head groups on the one side to that of the apposing side.

\begin{figure}[tb]
\includegraphics[clip,width=\linewidth]{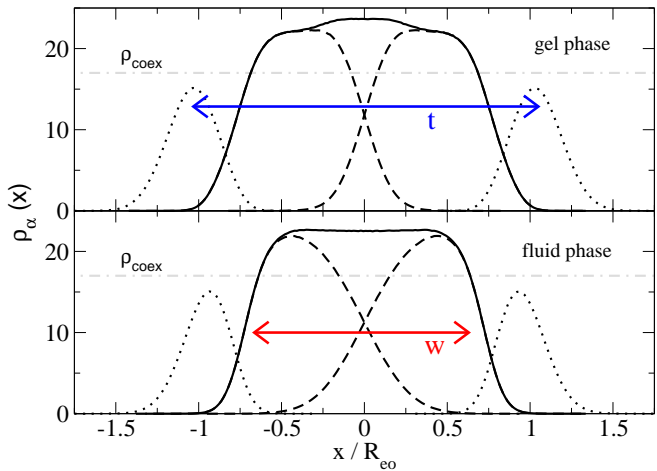}
\caption{\label{fig:profiles}
Molecular density profiles $\rho_\alpha(x)$ across the bilayer with respect to the local midplane at $\rho_{\text{coex}}=17$. The upper graph shows the density profile of the $L_\beta$ phase, while the lower one presents the $L_\alpha$ phase. The dotted lines denote the densities of the hydrophilic beads, the dashed lines indicate the densities of the hydrophobic beads for each leaflet, and the solid line marks the sum of the two hydrophobic densities. Finally, the dash-dotted line indicates $\rho_{\text{coex}}$, which has been used to derive the interactions of the hydrophobic beads.}
\end{figure}
Fig.~\ref{fig:profiles} depicts two density profiles, $\rho_\alpha(x)$, across the bilayer. They correspond to the liquid and gel phases, $L_\alpha$ and $L_\beta$, which have been observed at $\rho_{\text{coex}}=17$. Both show separated peaks for the hydrophilic heads and the hydrophobic tails, which implies that the coarse-grained lipids indeed form bilayer membranes. A closer inspection of the densities of the hydrophobic interior shows, that the two leaflets are clearly distinguishable, but that there is no dip in the center of the density profile, like it is known from atomistic or systematically coarse-grained models including solvent. In fact, we find a flat profile in the fluid phase and a hump in the gel phase, the latter being caused by an overlap of the last bead of the lipids from each side (not shown). 

The remaining overlap between the apposing leaflets in the liquid phase and the hump in the density profile in the gel state might arise from three different reasons: (i) The molecular shape in our model is rather finely discretized and the lipid tails are rather flexible. If the molecular shape becomes more rod-like, the density profile at the center is expected to develop a dip due to molecular packing. This could be achieved by a decrease of the number of beads per lipid or an increase of the bond stiffness. (ii) If the incompatibility between hydrophobic and hydrophilic segments increases, the bilayer thickness will increase and the interdigitation between the apposing leaflets will decrease. (iii) The flat density profile could also arise from the lack of solvent molecules. Since there is no solvent exerting pressure on the membrane, the lipids might have to interdigitate slightly, so that the whole bilayer remains stable. This would indicate a general problem in solvent-free models. We are not aware, however, of density profiles in the gel phase for another solvent-free model.

Although the width, $w$, of the hydrophobic core and the area per lipid, $\mv{a}=\mv{A}/(n/2)$, depend on the details of the soft, coarse-grained model, the dimensionless aspect ratio, $w / \sqrt{\mv{a}}$, can straightforwardly be compared to experiments. The most common two-tailed lipids have aspect ratios in the range of $w/\sqrt{\mv{a}} \approx 3-5$, whereas our simulations for single-tailed lipids yields $6 \leq w / \sqrt{\mv{a}} \leq 7$.  If one assumes that two single-tailed lipids of our coarse-grained model glued together result in one two-tailed lipid, one will double the mean area per lipid, $\mv{a}$, and obtain an additional factor of $\sqrt{2}$ in the denominator of the aspect ratio, thereby, obtaining aspect ratios that are in good agreement with experimental values. Noteworthy, it is impossible to obtain the right aspect ratio simply by selecting different interaction coefficients at fixed discretization, $N$, and fixed molecular architecture. At this level of coarse-graining the latter is important and must be taken into account, if one tries to map a specific kind of lipid. In this study, however, we are content with the simplest molecular architecture, i.e. linear molecules. We establish a conversion factor between the unit of length in the simulation and in experiments using the mean area per molecule. For synthetic lipids like DPPC, DPPE, or DLPC $\mv{a}\approx60~$\AA$^2$,\cite{Nagle96,Petrache00} so that we find in the fluid phase at $\rho_{\text{coex}}=17$ an equivalence of $1~\R=3.8~$nm.

\subsubsection{Elastic properties}
In the $NP_tT$-ensemble the projected area of the bilayer, $A$, is fluctuating. These fluctuations are related to the area compressibility, $k_A$, via \cite{Otter05,STEVENS04,LINDAHL00}
\begin{equation}
\label{k-a}
k_A = \beta \frac{\Mv{A^2}-\Mv{A}^2}{\Mv{A}}.
\end{equation}
This formula neglects undulations of the bilayer, which result in a difference between the projected and the true surface area of the bilayer.\cite{Otter05} However, for the small patches of a bilayer used in this study, this difference is negligible.

We have measured $k_A$ according to Eq.~\eqref{k-a} for the fluid and the gel phase at $\rho_{\text{coex}}=17$ as well as the fluid phase at $\rho_{\text{coex}}=40$. Using the conversion factor for the unit of length from above and converting the area compressibilities to area compression moduli, $K_A=k_A^{-1}$, we find $K_A=79.4~$mN/m for the fluid phase at $\rho_{\text{coex}}=17$, $67.2~$mN/m at $\rho_{\text{coex}}=40$, and $2045~$mN/m in the gel phase at $\rho_{\text{coex}}=17$. Typical experimental results for two-tailed lipids in the fluid phase yield values of $K_A \approx 240~$mN/m.\cite{Marsh06} We attribute the larger area fluctuations of our model bilayer to the softer interactions and the reduced number of degrees of freedom.

Another important quantity is the bending rigidity, $\kappa$, which measures the cost of undulations of the bilayer.\cite{Canham70,Helfrich73,Evans74} It is frequently calculated in particle-based simulations.\cite{Muller96,BOEK05,Brannigan06,STEVENS04,FARAGO03,LINDAHL00,Cooke05} If the fluctuations are small, the free energy of a curved membrane is given by the Helfrich-Hamiltonian
\begin{equation}
\label{helfrich}
\mathcal{H} = \frac{1}{2}\int {\rm d}^2r \, \left[\kappa \left(\nabla^2 h\right)^2
+ \Sigma \left(\nabla h\right)^2 \right],
\end{equation}
where $h(\mathbf r)$ in the Monge gauge is the bilayer's height above a reference plane. By inserting the Fourier expansion, $h(\mathbf r) = \sum_{\mathbf q} h_{\mathbf q} \exp\left({i \mathbf{qr}}\right)$ with $\mathbf q = 2\pi (n_x, n_y)/L$, in Eq.~\eqref{helfrich}, we observe that the different modes, $h_{\mathbf q}$, decouple, and, using the equipartition theorem, we obtain for the power spectrum \cite{Seifert97}
\begin{equation}
\label{helfrich-power-spectrum}
\Mv{\Abs{h_q}^2} = \frac{k_BT}{A\left(\kappa q^4 + \Sigma q^2\right)}.
\end{equation}
Eq.~\eqref{helfrich-power-spectrum} has been derived for the canonical ensemble, i.e., a constant box size. Here, we also used it to fit our data in the $NP_tT$-ensemble. Since the lateral box lengths fluctuate in this ensemble, $\mv{\abs{h_{\mathbf q}}^2}$ denotes the mean intensity of each mode at the time-averaged wave vector $\mathbf q = \mathbf n \cdot \mv{2\pi/L(t)}$. Additionally, we included the spectral damping factor in the calculation of $\mv{\abs{h_{\mathbf q}}^2}$ that arises from the interpolation of the continuous bilayer position onto a grid.\cite{Cooke05} From a computational point of view, we calculate the bilayer position, $h(\mathbf r)$, on a quadratic $16 \times 16$ grid by averaging over the perpendicular distances of all hydrophobic beads from the reference plane. Then we perform an FFT of $h(\mathbf r)$ yielding the amplitudes $h_{\mathbf q}$.

\begin{figure}[tb]
\includegraphics[clip,width=\linewidth]{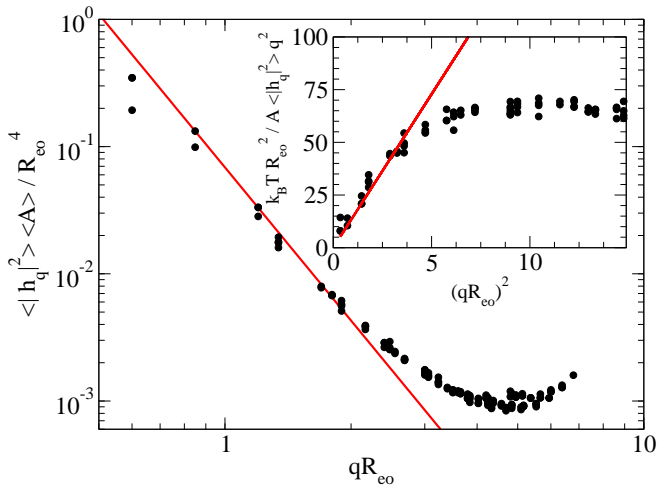}
\caption{\label{fig:bending-rigidity}
Power spectrum of the height fluctuations $\mv{\Abs{h_{\mathbf q}}^2}$ of the fluid phase at $\rho_{\text{coex}}=17$.  Inset: $k_BT / A \mv{\Abs{h_{\mathbf q}}^2} q^2$ plotted as a function of $q^2$. For small $q^2$ the data points are fitted to a straight line through the origin with slope $\kappa$.}
\end{figure}
Fig.~\ref{fig:bending-rigidity} shows the power spectrum of bilayer fluctuations in the fluid phase at $\rho_{\text{coex}}=17$ in a tensionless state. By fitting $k_BT / A \mv{\Abs{h_{\mathbf q}}^2} q^2$ as a function of $q^2$ to a straight line through the origin, we extract $\kappa$ from the slope (cf.~Eq.~\eqref{helfrich-power-spectrum}). Experimental values of the bending rigidity $\kappa$ lie for most biological membranes within a range of $5-60~k_BT$,\cite{Marsh06} and our results match this order of magnitude for both fluid systems under study. However, no bending rigidity could be obtained by this method in the gel phase.

Another independent, but rather crude estimate of $\kappa$ is provided by the area compressibility. It has been suggested that \cite{GOETZ99,LINDAHL00,Brannigan04,STEVENS04}
\begin{equation}
\label{kappa2}
\kappa \approx \frac{t^2}{b \cdot k_A},
\end{equation}
where $t$ is the thickness of the bilayer. There has been some debate about the value of the geometric factor $b$. Here we use $b=48$.
The obtained values match the order of magnitude of the values extracted from the undulation spectra.

\subsubsection{Diffusion}
Finally, we have measured the lateral mean-square displacements of the lipids' center of mass, and obtained the two-dimensional self-diffusion coefficient, $D$, from
\begin{equation}
\label{d-2d}
D = \lim_{t\to\infty} \frac{1}{4t} \Mv{\bigl(\mathbf r_i^{\text{cm}}(t)
- \mathbf r_i^{\text{cm}}(0)\bigr)^2}.
\end{equation}
This equation neglects all undulations but it is a good approximation for the systems under study.\cite{Reister05c,Reister07}

We used $D$ (see Tab.~\ref{tab:prop}) to establish a mapping between the unit of time, $\tau$, in the simulation and experimental units. Using $1~\R=3.8~$nm and a typical lipid diffusion coefficient at room temperature of $D=5~\mu$m$^2$/s,\cite{Kahya04} we obtain $1~\tau = 1.5~$ns. It is interesting to relate this identification of time scale in our coarse-grained model to the occurrence of flipflop events. Unfortunately, the flipflop rate hardly deviated from zero; we have observed only a very small number of events even in the longest simulation runs. Therefore only a lower bound for the mean time, $\mv{t}$, between two flipflop events is presented here. We find $\mv{t} \gg 10^8~\tau \approx 0.15 ~ s$. This is reasonable, since a passive flipflop is a thermally activated process, which happens on an experimental time scale of 1 event per molecule per day.\cite{Abreu04,Liu05}


\section{Main phase transition}
\label{main-phase-transition}
Depending on the control parameters, $\kappa N$ and $\rho_{\text{coex}}$, the lipids self-assembled into bilayers of different thermodynamic phases. We observe the fluid phase, $L_\alpha$, the non-interdigitated gel phase, $L_\beta$, the fully-interdigitated gel phase, $L_{\beta I}$, and a tilted gel phase, $L_{\beta '}$.\cite{Kranenburg05} Among the different phase transitions, the main phase transition, $L_\beta \leftrightarrow L_\alpha$, is definitely the most-important one. It has many of the characteristics well known from first-order transitions, like pronounced hysteresis effects, the occurrence of metastable states, and sharp peaks in the response functions.  

We used three different, but not independent methods to locate phase coexistence. First, we have applied a combination of Umbrella Sampling (US) and the Weighted Histogram Analysis Method (WHAM) to compute the free energy,\cite{umbrella,Ferrenberg89, WHAM,Souaille01,Chipot07} $F(\rho_{\text{coex}})$, in the vicinity of the main phase transition. Second, we have utilized Free Energy Perturbation theory (FEP) to extrapolate the free energy branches of each phase.\cite{Chipot07} Finally, we have used a histogram reweighting scheme to calculate the specific heat, $C(\rho_{\text{coex}})$.\cite{Ipsen90,Mouritsen83}


\subsection{Order parameters}
Several order parameters characterize the main phase transitions.\cite{Cooke05,Leekumjorn06,Revalee08} We chiefly employ the orientational order parameter
\begin{equation}
\label{order-parameter1}
S = \frac{1}{n}\Mv{\sum\limits_{i=1}^n\sum\limits_{j=0}^{N_A-1}
\frac{3\cos^2\theta_{j,j+1} - 1}{2(N_A-1)}},
\end{equation}
where
$\cos\theta_{j,j+1} = \mathbf n \cdot (\mathbf r_{j+1} - \mathbf r_j) /
\Abs{\mathbf r_{j+1} - \mathbf r_j}$
denotes the angle between the local, normal vector, $\mathbf n$, to the bilayer and the bond vector between two succeeding hydrophobic beads $j$ and $j+1$.  $i$ sums over all molecules in the bilayer and the average is taken over an ensemble of bilayers. $S=1$ means that all lipids are perfectly aligned parallel to $\mathbf n$, $S=0$ indicates isotropically distributed directions, and in the case $S=-1$ the lipids are perfectly aligned in the plane of the bilayer.  Since $\mathbf n(\mathbf r_i)$ is a function of the coordinates of many lipids, its calculation involved a triangulation procedure,\cite{Otter05} where we described the bilayer midplane by a set of small triangles with a unique normal, $\mathbf n(\mathbf r_i)$, in each triangle.

In the gel phase the lipids form a two-dimensional structure with $6$-fold symmetry. We probe this intermolecular packing by the order parameter
\begin{equation}
\label{order-parameter-psi6}
\psi_6 = \frac{1}{n}\Mv{\Abs{ \sum\limits_{i=1}^n \frac{1}{n_i}
\sum\limits_{j=1}^{n_i} \exp{\left(6i\phi_{ij}\right)} }}.
\end{equation}
Here, $n_i$ denotes the number of lipids adjacent to lipid $i$ (as determined by a Voronoi tesselation), and $\phi_{ij}$ the angle between the vector from the center of mass of lipid $i$ to that of lipid $j$, and some arbitrary but fixed direction in the plane of the bilayer. $\psi_6=1$ indicates perfect hexagonal symmetry over the entire bilayer, whereas $\psi_6=0$ signals the absence of bond-orientational order.

Both order parameters, $S$ and $\psi_6$, clearly distinguish between the liquid and the gel phase, but they differ in one crucial point: $S$ is composed of additive contributions, which only stem from conformational, single-molecule properties and therefore its change in response to moving a segment can be easily computed. The opposite is true for $\psi_6$, which only reflects bond-orientational order caused by intermolecular packing and requires the computationally intense Voronoi tessellation in order to identify the neighbors of a lipid. The main phase transition simultaneously involves both, a change in the in-plane degrees of freedom that dictate the bond-orientational order and a change in the conformational degrees of freedom.\cite{Mouritsen91}

We choose $S$ as the single reaction coordinate (order parameter) for the liquid-gel transition, because the conformational and the bond-orientational transition are coupled, and $S$ is considerably easier to compute than $\psi_6$.

\subsection{Determination of coexistence point}
Here, FEP has been used to calculate the free-energy difference between two systems that only differ in their non-bonded interactions, or more precisely, that only differed in the parameter, $\rho_{\text{coex}}$. To this end, we sample configurations at a reference density, $\rho_0$, and calculate the free energy difference, $F(\rho)-F(\rho_0)$, with respect to a system with a different density, $\rho$, by \cite{Chipot07}
\begin{equation}
\label{fep-formula}
F(\rho) - F(\rho_0) = - k_BT \ln
\Bigl\langle \exp{\bigl(-\beta
\Delta\mathcal{H}_{\text{nb}}(\rho)\bigr)} \Bigr\rangle_0.
\end{equation}
Here $\Mv{\cdots}_0$ stands for an ensemble average of the reference system and $\Delta\mathcal{H}_{\text{nb}}(\rho)$ is the difference of the non-bonded energies between these two systems,
\begin{eqnarray}
\nonumber
\Delta\mathcal{H}_{\text{nb}}(\rho) &=& \mathcal{H}_{\text{nb}}(\rho)
- \mathcal{H}_{\text{nb}}(\rho_0) \\
\nonumber
&=& \left[v_{\alpha\beta}(\rho)-v_{\alpha\beta}(\rho_0)\right]
\frac{P_{\alpha\beta}}{2} \\
\label{nb-diff}
&& + \left[w_{\alpha\beta\gamma}(\rho)-w_{\alpha\beta\gamma}(\rho_0)
\right]\frac{Q_{\alpha\beta\gamma}}{3}.
\end{eqnarray}
where the integrated densities, $P_{\alpha\beta}$ and $Q_{\alpha\beta\gamma}$ are defined by
\begin{eqnarray}
P_{\alpha\beta} &\equiv& \sum\limits_i \delta_{\alpha t(i)}
\tilde\rho_{2\beta}\left(\mathbf r_i\right) \\
Q_{\alpha\beta\gamma} &\equiv& \sum\limits_i \delta_{\alpha t(i)}
\tilde\rho_{3\beta}\left(\mathbf r_i\right)
\tilde\rho_{3\gamma}\left(\mathbf r_i\right).
\end{eqnarray}
Since FEP samples only the phase space of one thermodynamic phase, we cannot locate the coexistence of two phases. However, it is well suited to explore the free energy branch $F(\rho)$ of a single phase.

In contrast to FEP, the combination of US and WHAM allows a direct location of the phase coexistence. At first the free energy profile, $F(S)$, as a function of the order parameter, $S$, is calculated for a specific set of expansion coefficients. 
Let $S_{\text{fl}}$ and $S_{\text{gel}}$ denote the order parameter in the fluid and in the gel phase, respectively.  To obtain $F(S)$, bilayer configurations have to be uniformly sampled for all values of $S$ in the interval $S_{\text{fl}} \le S \le S_{\text{gel}}$. However, the unfavorable configurations in the miscibility gap are unreachable by conventional Boltzmann sampling because their statistical weight is exponentially small. By including an additional US potential, $W_i$, we force the system to sample also these unfavorable configurations. Specifically, we add the harmonic potential
\begin{equation}
\label{us-potential}
W_i = \frac{k}{2} n(N_A - 1) \left[S - S_i\right]^2.
\end{equation}
that biases the simulation to keep $S$ in the vicinity of $S_i$. Here $k=1~k_BT$ is a spring constant that measures how strong deviations from $S_i$ are penalized. We have used an equidistant spacing of the $S_i$ with $\Delta S_i = 0.01$ in the range $0.15 < S_i < 0.8$ to sample the whole interval uniformly.

A simulation has been performed for each value of $S_i$, in which we have recorded a trajectory of the order parameter, $S$, the total energy, $U$, and the integrated densities, $P_{\alpha\beta}$ and $Q_{\alpha\beta\gamma}$. These quantities are used to reweight the trajectories to different values of $\rho_{\text{coex}}$ (cf.~Eq.~\ref{nb-diff}). Each of them is binned into a normalized histogram that measures the biased probability density of visiting $S$ in a run with potential, $W_i$. In the subsequent weighted histogram calculation, this bias is removed from the histograms and all individual histograms are combined into one unbiased histogram, $p_0(S)$, in a way that the statistical error is minimal. For brevity, we omit the computational details and refer to the original work.\cite{WHAM,Souaille01} Once the Boltzmann probability distribution, $p_0(S)$, is available, the free energy, $F(S)/k_BT$, is computed as the negative logarithm.

$F(\rho,S)$ has been computed by histogram reweighting similar to Eq.~\eqref{fep-formula}. In contrast to the FEP calculations of the pure phases, $F(\rho,S)$ includes contributions from both phases. By taking the integral of $F(\rho,S)$ over all $S$ we obtained $F(\rho)$; a phase transition in this quantity is visible as a point with a rapidly varying derivative because finite size effects lead to a rounding of the transition.

We have also applied the reweighting procedure to $U$, so that the probability distribution $p(\rho, U)$ becomes available. From this quantity we calculate the mean total energy, $\mv{U}(\rho)$, and the specific heat 
\begin{equation}
\label{spec-heat-eq}
\frac{C}{k_B} = \frac{\left(\Mv{U^2}-\Mv{U}^2\right)}{(k_BT)^2},
\end{equation}
which serve to locate a first-order transition.

A first estimate of the position of the main phase transition is obtained from the center and the width of a hysteresis loop. Therefore we have simulated pre-assembled bilayers with 1600 lipids that were initially in the fluid phase at $\rho_{\text{coex}}=40$. We have performed several succeeding cycles with $\rho_{\text{coex}}$ running from $11$ to $40$ and vice versa in steps of $\Delta\rho_{\text{coex}}=0.5$ or $1.0$ for $\kappa N = 50,75,100,125$.

\begin{figure}[tb]
\includegraphics[clip,width=\linewidth]{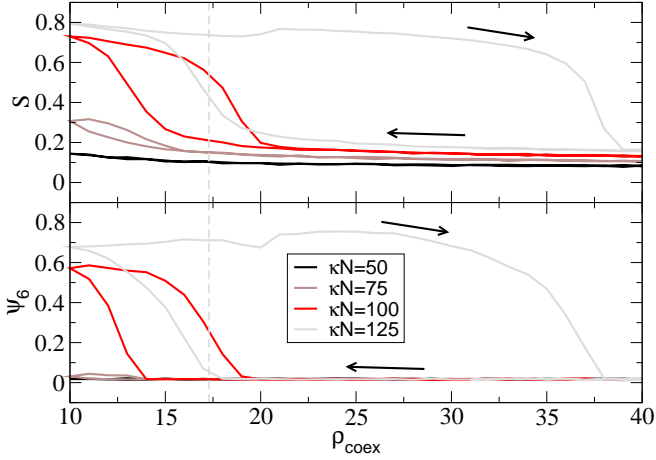}
\caption{\label{fig:hysteresis}
Hysteresis loops of the order parameters $S$ and $\psi_6$ for $\kappa N=50 \dots 125, \chi N=30$. At each step, the bilayer was simulated for $\Delta t = 100~\tau$. The step size between two simulations was $\Delta\rho_{\text{coex}}=1.0$, and the arrows mark the direction, in which $\rho_{\text{coex}}$ was proceeded. The dashed line at $\rho_{\text{coex}}=17.27$ indicates the transition point for $\kappa N=100$.}
\end{figure}
Near the main phase transition, large hysteresis effects occur in $S$ and $\psi_6$ as shown in Fig.~\ref{fig:hysteresis}. The loops for both order parameters differ only quantitatively. They are weakly shifted, and intramolecular order persists up to slightly higher $\rho_{\text{coex}}$ than the intermolecular order. The widths of the loops grows with increasing $\kappa N$, i.e., metastable domains persist up to higher $\rho_{\text{coex}}$. Additionally, the amplitudes of the order parameters increase indicating different thermodynamic phases. For instance, in the case $\kappa N=125$, two distinct gel phases ($L_\beta$ and $L_{\beta'}$) occur. Their transition is visible as a dip in both order parameters near $\rho_{\text{coex}}\approx 20$.  


%

We focus on the system, $\kappa N=100$ and $\chi N=30$, where the results in Fig.~\ref{fig:hysteresis} have indicated that the main phase transition is located in the interval $15 < \rho_{\text{coex}} < 18$. We calculate $F(S)$ by means of US/WHAM from simulations at different coexistence densities, $\rho_{\text{coex}}=16.28, 17.00, 17.29$, with $2-5$ different initial configurations in the $NP_tT$-ensemble for $\Delta t = 3000~\tau$.  It is advantageous to start the simulation from an initial configuration where both phases are already present.\cite{comment1} 

\begin{figure}[tb]
\includegraphics[clip,width=\linewidth]{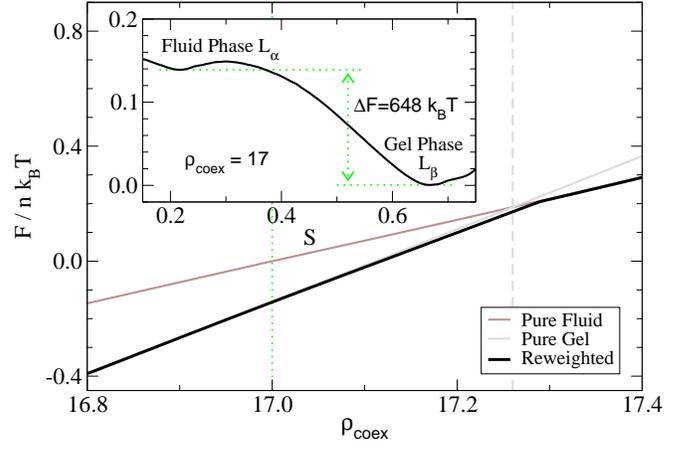}
\caption{\label{fig:fep-reweighting}
Inset: Free energy, $F(S)$, at $\rho_{\text{coex}}=17$. There was a difference of $\Delta F=648~k_BT$ between the minima of the $L_\alpha$ phase and the $L_\beta$ phase. Main Panel: $F(\rho_{\text{coex}})$ obtained from histogram reweighting in comparison to two independent FEP calculations of a pure $L_\alpha$ and a pure $L_\beta$ phase, whose offset $\Delta F$ at $\rho_{\text{coex}}=17$ was known from the inset. Both curves intersected at $\rho_{\text{FEP}}=17.26$ (dashed gray line).} 
\end{figure}
The inset of Fig.~\ref{fig:fep-reweighting} shows $F(S)$ at $\rho_{\text{coex}}=17$. The two visible minima correspond to the metastable $L_\alpha$ and the stable $L_\beta$ phase.  However, the offset $\Delta F = 648~k_BT$ between these minima indicates that the gel phase is thermodynamically stable. To locate the phase transition, we reweight $\rho_{\text{coex}}$ searching for a rapid variation (i.e., rounded discontinuity) of the slope that signals the phase transition. Such a kink occurs at $\rho_{\text{US}}=17.29$ (cf.~main panel of Fig.~\ref{fig:fep-reweighting}) indicating the crossing of the free energy branches of the different phases. 

\begin{figure}[tb]
\includegraphics[clip,width=\linewidth]{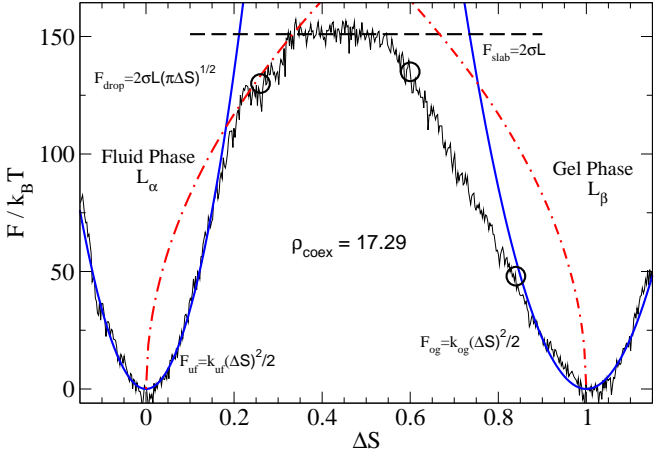}
\caption{\label{fig:free-energy-normalized}
$F(\Delta S)$ at phase coexistence compared to the phenomenological expressions from eqs.~\eqref{f-sv}, \eqref{f-drop}, and \eqref{f-slab}.  Circles mark the values of $\Delta S$, at which typical configurations are visualized in Fig.~\ref{fig:droplet-slab}.}
\end{figure}
It is convenient to introduce a normalized order parameter, $\Delta S \equiv (S - S_{\text{fl}}) / (S_{\text{gel}} - S_{\text{fl}})$, so that the minima of the free energy in the fluid phase at $S_{\text{fl}}$ and in the gel phase at $S_{\text{gel}}$ correspond to $\Delta S=0$ and $\Delta S=1$, respectively. Fig.~\ref{fig:free-energy-normalized} depicts $F(\Delta S)$ at $\rho_{\text{coex}}=17.29$. At this point both phases have equal statistical weight, and they are separated by a free energy barrier with a plateau value of $F_{\text{slab}}=151.0(5)~k_BT$. We calculate $\Delta S$ from the abscissae of the minima, $S_{\text{fl}}=0.212$ and $S_{\text{gel}}=0.625$.

\begin{figure}[tb]
\includegraphics[clip,width=\linewidth]{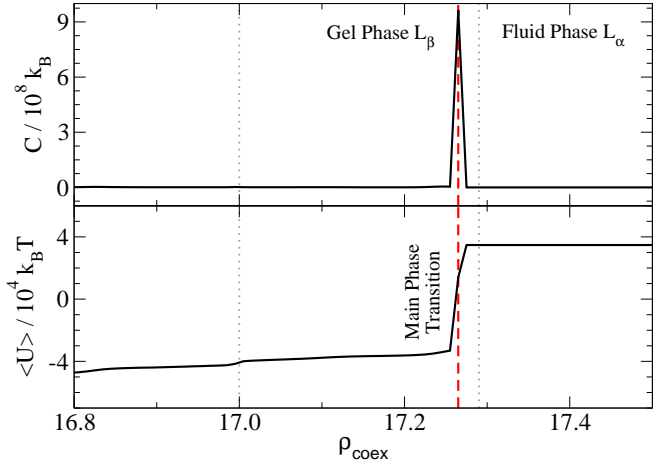}
\caption{\label{fig:specific-heat}
Specific heat $C$ (above) and total energy $\mv{U}$ (below) obtained from histogram reweighting as a function of $\rho_{\text{coex}}$. The main phase transition is visible as a sharp rise in $\mv{U}$, but also as a peak in $C$ at $\rho_{\text{SH}}=17.27$ (dashed red line).  The dotted gray lines mark the points where simulations using US have been performed.}
\end{figure}
To confirm the transition point, we employ Eq.~\eqref{spec-heat-eq} to calculate $C(\rho_{\text{coex}})$ and $\mv{U}(\rho_{\text{coex}})$ by reweighting (see Fig.~\ref{fig:specific-heat}). The main phase transition is visible as a sharp peak in $C(\rho_{\text{coex}})$ at $\rho_{\text{SH}}=17.27$, as well as a steep rise in $\mv{U}(\rho_{\text{coex}})$ at the same position. The slow rise of $\mv{U}$ in the interval $16.8 < \rho_{\text{coex}} < 17.27$ can be attributed to the gradual melting of the hydrophobic tails.

Finally we have conducted two additional, independent simulations at $\rho_{\text{coex}}=17$ in the $NP_tT$-ensemble without an US potential. One initial configuration was prepared in a pure fluid phase ($L_\alpha$, metastable) and the other was prepared in a pure gel phase ($L_\beta$). The free energy branches of each phase are extrapolated with FEP (cf.~Eq.~\eqref{fep-formula}). In this method the relative free energy difference between both branches remains undetermined.  However, it has already been computed by the offset between the branches at $\rho_{\text{coex}}=17$ yielding $\Delta F = 648~k_BT$. The main panel of Fig.~\ref{fig:fep-reweighting} depicts the two, correspondingly shifted branches of $F(\rho_{\text{coex}})$, which intersected at $\rho_{\text{FEP}}=17.26$.

Gratifyingly the US/WHAM results for the free energy $F(\Delta S)$ and $F(\rho_{\text{coex}})$ are consistent, indicating the high statistical accuracy of our data. In Fig.~\ref{fig:fep-reweighting} we present the two branches of the free energy, $F(\rho_{\text{coex}})$, of each phase obtained from FEP in comparison to the result from US/WHAM. In the fluid phase both methods completely agree, however, in the gel phase there is a small difference discernable. Between $\rho_{\text{coex}}=17.15$ and $17.29$ the FEP calculation slightly overestimates the free energy by $\Delta F=0.02~k_BT$ per lipid, which arise from a gradual loss of bond-orientational order as the transition is approached from the gel phase. Therefore the result, $\rho_{\text{FEP}}$, is less accurate than the other estimates.

A similar way of determining the phase coexistence point has been applied earlier.\cite{Mouritsen83} In that study the relationship between the branches is fixed by the knowledge of the two bulk free energies, which were extracted from mean-field theory. A similar calculation is possible in the fluid phase of our coarse-grained model, but it is not accurate in the gel phase, where correlations between the lipids are essential. These correlations, which are captured by in our simulations, are clearly visible, e.g., in the order-parameter $\psi_6$.

The three estimates of the location of phase coexistence, $\rho_{\text{US}}$, $\rho_{\text{FEP}}$, and $\rho_{\text{SH}}$, nicely agree with each other. The main error source of our estimate of the phase coexistence, however, stems from possible sampling error along the US path that reversibly connects the liquid and the gel phase, which are difficult to estimate. The consistency of the results suggests that the liquid-gel transition for the parameters, $\kappa N=100$ and  $\chi N=30$ , occurs at
\begin{equation}
\rho^{*}_{\text{coex}}=17.3(1).
\end{equation} 
Thus, the uncertainty in the location of the transition point is reduced by a factor of 50 compared to the uncertainty in the hysteresis loops.

\subsection{Bilayer configurations}
Besides the point of the phase coexistence, $F(\Delta S)$ also offers an insight how typical configurations of the finite bilayer inside the miscibility gap look like.\cite{Binder82,Berg93,Hunter95,Binder03c,MacDowell04} Let us consider a lipid bilayer in the fluid phase. If $\Delta S$ is slightly increased from the value $\Delta S=0$ this small increase will be distributed homogeneously throughout the bilayer.  The excess free energy of this undercooled fluid bilayer up to second order in $\Delta S$ is given by a Taylor expansion around the minimum
\begin{equation}
\label{f-sv}
F_{\text{uf}} = \frac{k_{\text{uf}}}{2}\left(\Delta S\right)^2,
\end{equation}
where $k_{\text{uf}}$ is a constant measuring the response of the system to changes in $\Delta S$.

\begin{figure*}[tb]
\includegraphics[clip,width=\linewidth]{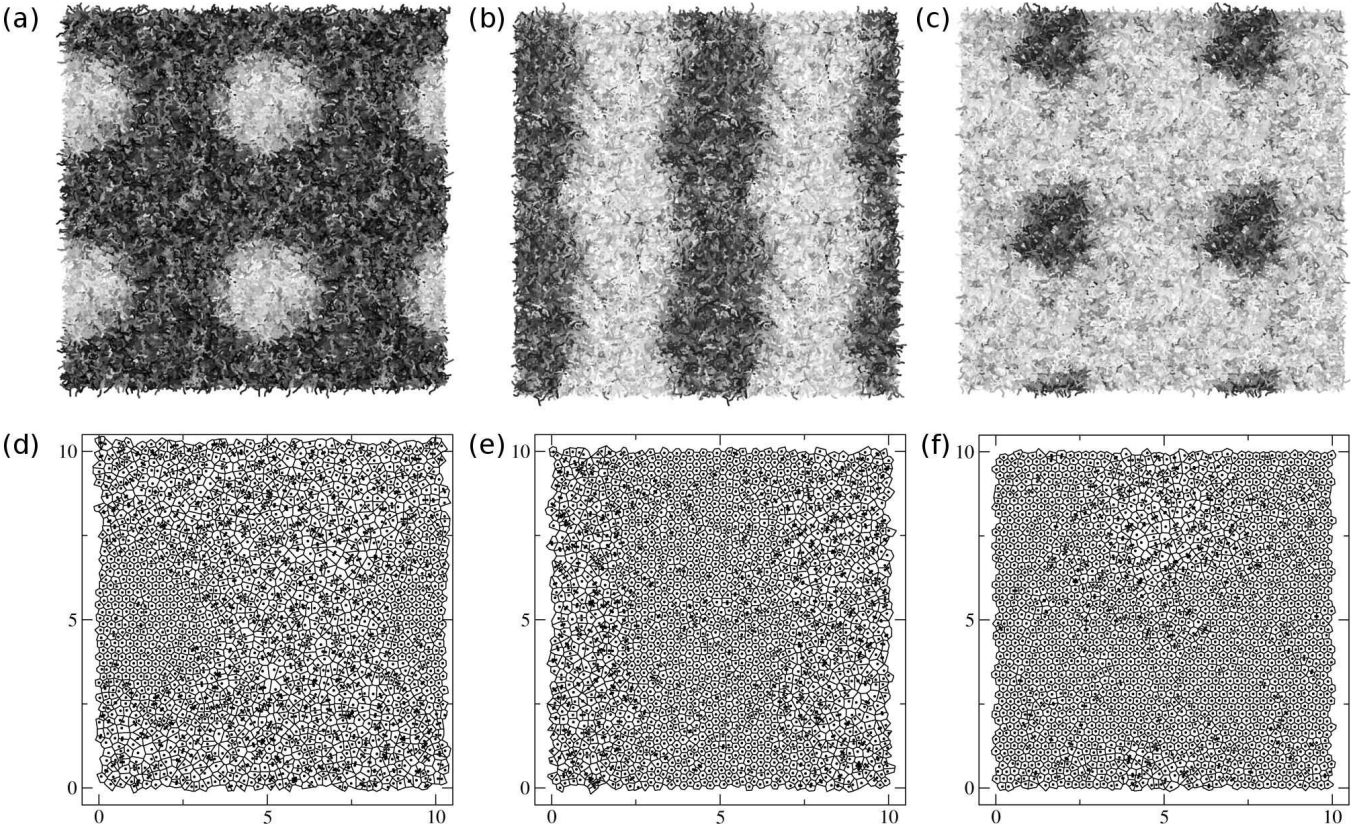}
\caption{\label{fig:droplet-slab}Typical bilayer configurations inside the miscibility gap (duplicated across each periodic boundary). (a) shows gel droplets in a fluid phase ($\Delta S=0.26$), (b) shows the slab geometry ($\Delta S=0.60$), and (c) shows fluid droplets in a gel phase ($\Delta S = 0.84$). Each lipid is colored by its local order parameter: the $L_\beta$ phase is white, the $L_\alpha$ phase is dark, and intermediate values are gray. Created with VMD.\cite{VMD} (d-f): Voronoi tesselation\cite{Shinoda97} of the same configurations as before, showing the area occupied by each lipid. The lipids in the $L_\beta$ phase ordered in a hexagonal structure, while no such structure is present in the $L_\alpha$ phase.}
\end{figure*}
In a macroscopic system an undercooled bilayer is metastable and the lipids will condense into two-dimensional droplets of radius, $R$, that consist of the
thermodynamically stable, gel phase (cf.~Fig.~\ref{fig:droplet-slab}a, c). In the framework of classical nucleation theory, the excess free energy of such a
droplet is given by the droplet's perimeter and the thermodynamic line tension, $\sigma$, i.e.  
\begin{equation}
\label{f-drop1}
F_{\text{drop}} = 2\pi\sigma R.
\end{equation}
Note, however, that the thermodynamic line tension depends on the length scale, i.e., the perimeter of the drop. Since the fluid and the gel phase have both the same free energy at coexistence, there is no bulk contribution to Eq.~\eqref{f-drop1} from the interior of the droplet. Since the lipids occupy in both phases roughly the same area (cf. Tab.~\ref{tab:prop}), the area of the droplet is in good approximation proportional to $\Delta S$, i.e.
\begin{equation}
\label{f-drop2}
\pi R^2 \simeq L^2 \Delta S.
\end{equation}
where we have used that the normalized order-parameter, $\Delta S$, quantifies the fractional area of the gel phase.
Combining Eqs.~\eqref{f-drop1} and \eqref{f-drop2}, one obtains $R \sim \Delta S^{1/2}$ and
\begin{equation}
\label{f-drop}
F_{\text{drop}} = 2\sigma L \sqrt{\pi \Delta S}.
\end{equation}
If the two-dimensional droplet grows larger, its size will become comparable to the linear dimension, $L$, of the simulation box. Then it is more favorable to form a gel phase slab that is separated from the fluid phase by two plane interfaces of length $L$ (cf.~Fig.~\ref{fig:droplet-slab}b).  In this case, the excess free energy is independent of $\Delta S$, i.e.
\begin{equation}
\label{f-slab}
F_{\text{slab}} = 2\sigma L
\end{equation}
Increasing $\Delta S$ even further, one observes the reverse set of configurations. The slab of the fluid phase grows thinner and thinner, and at some point it becomes favorable to form a fluid droplet surrounded by the gel phase. The radius of the fluid droplet decreases while $\Delta S$ increases.  Finally, the droplet vanishes and the lipids form a homogeneous gel phase.

A fit of $F(\Delta S)$ to Eq.~\eqref{f-sv} in the vicinity of the minima yields the two constants, $k_{\text{uf}}=6750~k_BT$ and $k_{\text{og}}=4290~k_BT$, that quantify the response of the bulk phases to changes in $\Delta S$. We indicate the resulting parabolas in Fig.~\ref{fig:free-energy-normalized}. In addition, we also plot $\Delta F_{\text{slab}}$ from Eq.~\eqref{f-slab}, as well as Eq.~\eqref{f-drop} for the droplet shape on each side of the free energy profile.

\begin{figure}[t]
\includegraphics[clip,width=\linewidth]{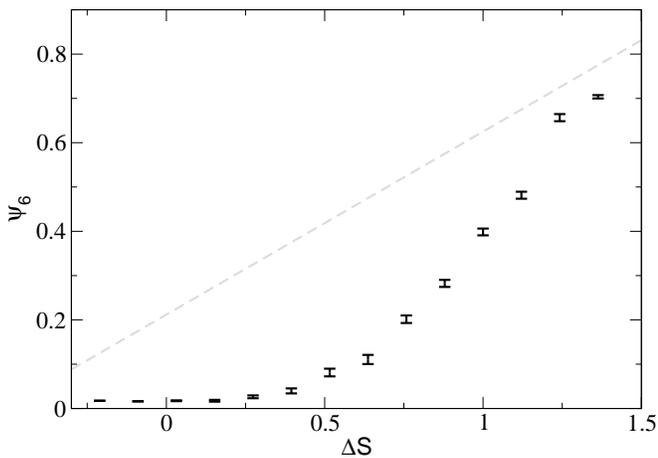}
\caption{\label{fig:s-psi6}
For $\Delta S<0.5$ only a small amount of the lipids had a straight conformation, and it was unfavorable to form gel domains; no hexagonal symmetry, measured by $\psi_6$, was visible. For $\Delta S>0.5$ enough lipids acquired such a conformation, so that gel domains formed and hexagonal symmetry set in quickly.}
\end{figure}
It is interesting to note that the left half of Fig.~\ref{fig:free-energy-normalized} with $\Delta S < 0.5$ fits the phenomenological expressions, Eqs.~\eqref{f-drop} and \eqref{f-slab}, well, while the data for higher values of the order parameter exhibit larger deviations. In addition, $k_{\text{uf}} > k_{\text{og}}$, i.e., the gel phase, $L_\beta$, has a smaller response with respect to changes in $\Delta S$ than the fluid phase, $L_\alpha$. This discrepancy stems from the onset of bond-orientational order. Fig.~\ref{fig:s-psi6} shows that bond-orientational order is weak for $\Delta S \lesssim 0.4$. For larger values of the order parameter, however, the lipids acquired also bond-orientational order, which results in a decrease of $F(\Delta S)$. The minimal free energy is finally reached in a state with hexagonal symmetry of the lipids. Only the trailing end beads of the lipids interdigitate with the ones from the apposing leaflet and constitute a thin, disordered layer at the center of the bilayer.

Additional deviations from the simple phenomenological estimates arise from the interaction between the lines that separate the liquid and the gel domains. For instance, the thermal fluctuations of the two lines in the slab geometry induce attractive Casimir forces.\cite{Dean07} While the slab is growing thinner and thinner at the crossover to the droplet geometry, these Casimir forces become more pronounced and reduce the free energy. However, there is an additional reduction of the free energy coming from transversal line fluctuations.\cite{Golestanian96} These become pronounced for small widths of the slab and finally lead to its destruction. While it is in principle possible to study these fluctuation mediated interactions by careful inspection of $F(\Delta S)$ at the edges of the plateau, we did not investigate these effects in further detail. 


\section{Line tension}
\label{line-tension}
In this section we present two different methods for obtaining the line tension from the slab-configurations in the middle of the miscibility gap, extracting the bare line tension, $\lambda$, and the thermodynamic line tension, $\sigma$.

\subsection{Bare line tension, $\lambda$}
If the boundary line separating the two domains in the slab geometry is smooth and free of overhangs, one can describe its position by a function, $h(z)$. The statistical properties of $h(z)$ follow from the capillary wave Hamiltonian\cite{Gelfand90,Safran94}
\begin{equation}
\label{capillary-wave-hamiltonian}
\mathcal{H}_{\text{cap}} = \lambda L_z +
\frac{\lambda}{2} \int\limits_0^{L_z} {\rm d}z \, \left[h'(z)\right]^2,
\end{equation}
where $L_z$ is the projected length of the line.

Routinely, $\lambda$ is determined from $h(z)$. To this end, one expands $h(z)$ in a Fourier series and investigates the power spectrum of fluctuations, $\mv{\abs{h_q}^2}$. A fit to the expression
\begin{equation}
\label{line-spectrum}
\mv{\Abs{h_q}^2} = \frac{k_BT}{\lambda L_z q^2},
\end{equation}
which one derives from Eq.~\eqref{capillary-wave-hamiltonian} using the equipartition theorem (cf.~Appendix~\ref{capillary-wave-appendix}), then
yields $\lambda$.

\begin{figure}[tb]
\includegraphics[clip,width=\linewidth]{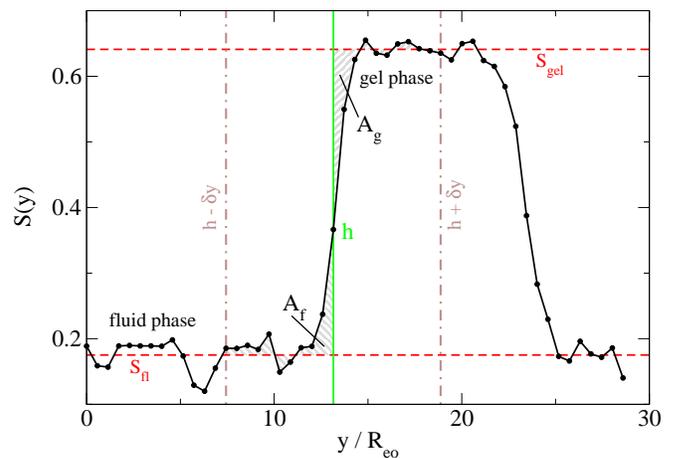}
\caption{\label{fig:gibbs-schema}
To calculate the Gibbs dividing surface the bulk values of the order parameter $S_{\text{gel}}$ and $S_{\text{fl}}$ are computed (dashed lines).  After that, the integral criterion is applied in a narrow interval $h-\delta y \cdots h+\delta y$ surrounding each edge of the slab.  The position of the interface $h$ is chosen, such that $A_f = A_g$.} 
\end{figure}
Several different schemes to locate the position of such an interface are known.\cite{chacon05} We use an integral criterion,\cite{Pastorino09,Werner99b} in which we subdivide the profile of the local order parameter, $S(y,z)$, into $N_z=16$ horizontal stripes with a width of $\Delta z = L_z / N_z$. Each stripe is binned into two histograms, one for each leaflet, with a bin width of $\Delta y = 2~\Delta L$ (where $\Delta L$ denotes the range of the non-bonded interactions), yielding in total $2N_z$ histograms per snapshot. The bulk values of the order parameter in the gel phase, $S_{\text{gel}}$, and in the fluid phase, $S_{\text{fl}}$, have been extracted once for each snapshot. As illustrated in Fig.~\ref{fig:gibbs-schema}, $h(z_i)$ in stripe $i$ has been calculated for each side of the slab separately as the position of the Gibbs dividing surface, such that
\begin{equation}
\int\limits_{h-\delta y}^{h}{\rm d}y ~ \bigl( S(y, z_i) - S_{\text{fl}}\bigr)
= \int\limits^{h+\delta y}_{h}{\rm d}y ~ \bigl(S_{\text{gel}} - S(y, z_i)\bigr).
\end{equation}
Here $\delta y=10~\Delta L$ is used to confine the integral to a narrow region surrounding the Gibbs dividing surface, so that fluctuations of the bulk influence the position of the interface minimally. Once the function, $h(z)$, is computed, the fluctuation power spectrum $\mv{\abs{h_q}^2}$ is calculated via FFT and averaged.

We compared two different ways of calculating this average. On the one hand, we argue that the bilayer is essentially a two-dimensional object, so that only the boundary lines at different sides of the slab, but not on different leaflets, fluctuate independently. This leads to an average calculated over two independent line configurations per snapshot. On the other hand, we can also locate the boundary in each leaflet independently and study their fluctuation spectra. Hence, the average included four different lines per snapshot. Finally, each average was divided by the spectral damping factor $\left[\sin(\pi n/N_z)/(\pi n / N_z)\right]^2$.\cite{Cooke05} For small lateral wave vectors, we expect that the boundaries in the two apposing leaflets are coupled and both methods yield the same, bare line tension in the limit, $q \to 0$. 
To record the fluctuation spectrum of the interface line a rectangular shape of the simulation box is chosen, with $L_y > L_z$. In this asymmetric situation, the slab will attain the lowest interfacial free energy if it aligns parallel to the $z$-axis. So the orientation of the interfaces is dictated by the system geometry. Specifically, we have assembled an initial configuration with $n=14040$ lipids, at $\rho_{\text{coex}}=17$ having $L_y=3L_z \approx 30~\R$.
Note that the length of the interface, $L_z$, is the same as in section~\ref{main-phase-transition}, only the other box length, $L_y$, has changed. This configuration is simulated in the $NP_tT$-ensemble with two independent degrees of freedom, $L_y$ and $L_z$, allowing anisotropic area fluctuations. Additionally, we employ a US potential with $S_0=0.34, k=5.0~k_BT$, driving the system to the desired slab geometry. After an equilibration time of $\Delta t=10^4 ~ \tau$ the box lengths fluctuate around mean values of $\mv{L_y} = 31.78~\R$ and $\mv{L_z} = 10.0~\R$, and both interfaces on both leaflets of the bilayer are flat. Unlike the experimental situation, where only domains of spherical shape are observed, there is no Laplace pressure because the interfaces are not curved and thus the two phases coexist at the same vanishing lateral pressure.

\begin{figure}[tb]
\includegraphics[clip,width=\linewidth]{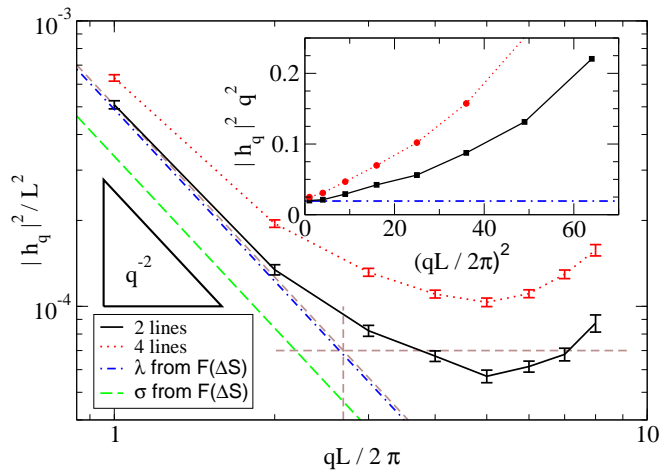}
\caption{\label{fig:line-spectrum}
Power spectrum $\Mv{\Abs{h_q}^2}$ of the boundary line fluctuations calculated for 2 (solid) and 4 (dotted) independent interface lines with the standard mean error in comparison to the value calculated from $F(\Delta S)$ (dash-dotted). The dashed grey lines indicate the estimated value of the UV-cutoff $q_{\text{max}}=2\pi/a$. Inset: $\Mv{\Abs{h_q}^2} q^2$ plotted as a function of $qL_z$. For small $q$ this function became constant and intersected the y-axis at $k_BT / \lambda L_z$.} 
\end{figure}
Fig.~\ref{fig:line-spectrum} shows the line fluctuation spectra for the two different ways of averaging with two and four independent lines. In the inset of Fig.~\ref{fig:line-spectrum}, $\mv{\Abs{h_q}^2} q^2$ is plotted as function of $(qL)^2$. For small $q$ this expression becomes linear and intercepts the y-axis at $k_BT / \lambda L_z$. Fits yield $\lambda_2=5.17~k_BT/\R$ for two lines and $\lambda_4=4.35~k_BT/\R$ for four lines, respectively.

For small wave vectors, $q$, both graphs in the fluctuation power spectrum of Fig.~\ref{fig:line-spectrum} asymptotically approach the same $q^{-2}$ power-law. This indicated that interface fluctuations are correlated on large scales. At larger values of $q$, the lines fluctuate independently and, concomitantly, the high-$q$ estimate of the line tension is lower for the data extracted from four lines than the data for two lines. For $qL_z/2\pi > 4$ the graph shows clear deviations from the simple power-law, which is expected, because the description of the line by the capillary wave Hamiltonian breaks down on microscopic scales. Thus, $L_z\approx 10~\R$ is too small to observe the expected $q^{-2}$--scaling over an extended $q$-range. We compute the final result by taking the average of both values $\lambda_2$ and $\lambda_4$ and the deviation as the error bar. This yields the final estimate of the bare line tension, $\lambda = 4.8(4)~k_BT/\R$.

\subsection{Thermodynamic line tension, $\sigma$}

A different measure of the free energy cost of a phase boundary -- the thermodynamic line tension, $\sigma$ -- is extracted from the free energy profile $F(\Delta S)$ in Fig.~\ref{fig:free-energy-normalized}. The excess free energy of the slab configuration is dominated by the interfacial free energy \cite{Binder82,MM_JUAN}, $F_{\text{slab}} = 2\sigma L_z$, provided that the system is large enough, for the two interfaces not to interact.  $F_{\text{slab}}$ is readily obtained from the plateau value of $F$ in the center of the free energy profile. However, there is a difference between the bare line tension, $\lambda$, and the thermodynamic line tension, $\sigma$, which depends on the length scale. It is shown in Appendix~\ref{capillary-wave-appendix} that the two quantities are related via
\begin{equation}
\label{thd-line-tension}
\sigma = \lambda + n_{\text{max}} \ln\left(2\pi\beta\lambda L_z\right)
+ 2\ln\left(n_{\text{max}}!\right),
\end{equation}
which describes the renormalization of the bare line tension $\lambda$ by fluctuations. $n_{\text{max}}$ denotes the closest integer to $L_z/a$ and $a$ is a UV-cutoff that characterizes the smallest scale, on which the fluctuations of the line are describable by a capillary-wave Hamiltonian. $a$ has to be independently determined.

To calculate the bare line tension, $\lambda$, from the height of the free energy barrier, we estimate the value of the UV-cutoff $a$ graphically from the intersect of the $q^{-2}$ power-law at small $q$ and a constant fluctuation strength at higher $q$. We find $q_{\text{max}}=2\pi/a=1.69 ~ \R^{-1}$, i.e., $a \approx 3.7~\R$ and thus $n_{\text{max}} = 3$.  By taking the plateau value $F_{\text{slab}} = 151.0(5)~k_BT$ from Fig.~\ref{fig:free-energy-normalized} we obtain $\sigma=7.55(3)~k_BT/\R$, and by solving Eq.~\eqref{thd-line-tension} numerically for $\lambda$, we obtain $\lambda_F = 5.45(3)~k_BT / \R$, which is in good agreement with the estimate obtained from the spectrum, $\lambda = 4.8(4)~k_BT/\R$.

To illustrate the difference between $\sigma$ and $\lambda_F$, we include in Fig.~\ref{fig:line-spectrum} the two asymptotical power spectra arising from $\lambda_F$ as the dash-dotted line, and from $\sigma$ when simply inserted into Eq.~\eqref{line-spectrum} as the dashed line. $\lambda_F$ nicely describes the measured fluctuation spectrum, whereas $\sigma$ results in a significantly damped spectrum. Thus, there is a notable difference between $\sigma$ and $\lambda_F$ of approximately $30~\%$ even for the small system size considered in the simulation.

Finally, we note that the line tensions, $\lambda$, of membranes formed by biologically relevant lipids are typically on the order of $10~$pN. \cite{Joannis06,Karatekin03,Esposito07} Using the previously determined scale factors for length and energy, we find that $1~\text{pN}\simeq 0.5~k_BT/\R$, i.e. both simulation results $\bar\lambda\approx 10(1)~$pN and $\lambda_F\approx 11(1)~$pN match the experimental values.


\section{Conclusions}
\label{conclusions}
In this study, we have presented a soft, solvent-free coarse-grained model for the simulation of lipid bilayers. The non-bonded interactions are inspired by a field-theoretic description and take the form of a third-order expansion of the excess free energy functional in the densities of the hydrophilic and hydrophobic beads. The numerical values of the expansion coefficients are related to a few thermodynamic key characteristics like the density of the hydrophobic core and its compressibility.\cite{KADV} The local structural properties of the bilayer, i.e., liquid-like packing of the coarse-grained beads, can be independently adjusted by means of weighting functions for the attractive pair-wise interactions and the repulsive triple interactions. In this way, we devise a flexible model where the interaction parameters bear a clear physical interpretation. With the described DPD simulation technique, it is possible to simulate bilayer patches of a size of $A=100 \times 100~$nm$^2$ for up to $\Delta t = 1~$ms.

Depending on the molecular asymmetry, the lipids self-assemble into spherical and cylindrical micelles, wormlike micelles, bilayers, and inverted structures. Static and dynamic properties of the bilayer membranes have been calculated. In the fluid phase, we obtained a bending rigidity of $\kappa\approx 15-19~k_BT$, an area compression modulus of $k_A \approx 70~$mN/m, and a molecular aspect ratio of $1:7$. These numbers match the orders of magnitude observed in experiments.

In the second part, the phase behavior of the bilayer has been investigated. Depending on the harshness of the short-range repulsive interactions and on the density of the hydrophobic interior, we observe the fluid phase, $L_\alpha$, and three different gel phases. We have studied the main phase transition, $L_\beta \leftrightarrow L_\alpha$, in detail. By means of Umbrella Sampling and the Weighted Histogram Analysis Method, we have calculated the free energy as a function of a conformational order parameter near the phase coexistence and obtained the free energy profile across the miscibility gap.

The phase coexistence has been accurately located by three different, although not independent, methods giving consistent results: (i) histogram reweighting, (ii) free energy perturbation calculations, and (iii) the calculation of the specific heat. Different geometries of the minority phase, like droplets and slabs, have been observed in the miscibility gap. Finally, the line tension separating different domains has been calculated by means of two different methods, giving a final result of $\lambda \approx 10~$pN, which is in excellent agreement with experimental studies.
The computational method outline in the present work is not restricted to studying phase transitions of single-component lipid membranes. For instance, one could study fluid-fluid coexistence in lipid mixtures or asymmetric bilayers. In such a case, one will setup simulations at constant lateral tension in a semigrand canonical ensemble, where a suitable order parameter will be the overall lipid composition. 

We hope that the soft, coarse-grained model for lipid bilayer membranes and the computational techniques will find further applications in the study of collective phenomena in membranes.

\begin{acknowledgments}
We have benefitted from many valuable discussions with K.Ch.~Daoulas and C.~Pastorino. Helpful comments from S.~Frank, G.~Marelli, Y.~Norizoe and J.~Shillcock are also acknowledged. M.H. thanks the DAAD for the support of a visit to Buenos Aires, Argentina. Financial support by the Volks\-wa\-gen foundation and the SFB 803 (TP B3) are gratefully acknowledged. Computing time was generously provided by the HLRN Hannover, the J\"ulich Supercomputing Centre (JSC), and the GWDG G\"ottingen.
\end{acknowledgments}

\appendix

\section{Integration algorithm}
\label{integrator}
Most of the simulations have been performed in a statistical ensemble, in which the average tangential pressure $\overline{P_t}$ and the height $L_x$ of the simulation box are kept constant, so that the area $A(t)$ enters as one additional dynamic degree of freedom.  Although several related algorithms have been published,\cite{Jakobsen05,Zhang95} we describe our symplectic integration algorithms for completeness in this appendix. It is derived from the Langevin piston method developed in Ref.~\cite{Kolb99} A second integration algorithm with two additional degrees of freedom ($L_y(t), L_z(t)$) can be derived in a similar way. It is used to simulate bilayers in the gel phase, where isotropic fluctuations of the area are inappropriate due to the hexagonal ordering of the lipids. For brevity the details of this similar algorithm are omitted.

In a thermally isolated system the first law of thermodynamics reads
\begin{eqnarray*}
{\rm d}E &=& - P {\rm d}V + \gamma {\rm d}A = -P {\rm d}V + L_x(P - \overline{P_t}) {\rm d}A \\
&=& - \overline{P_t} L_x  {\rm d}A,
\end{eqnarray*}
where $P$ is the normal pressure and $\gamma = L_x(P - \overline{P_t})$ is the surface tension. Hence, ${\rm d}H \equiv {\rm d}(E + \overline{P_t}L_xA) = 0$ and the enthalpy $H=E + \overline{P_t}L_xA$ is a conserved quantity. Following \citet{Anderson80} we now introduce scaled coordinates $y_i=s_{iy}\sqrt{A}$ and $z_i=s_{iz}\sqrt{A}$ tangential to the plane, retaining the normal coordinates, $x_i$. In the particle velocities, $\dot y_i=\dot s_{iy}\sqrt{A}$ the second term is deliberately omitted to achieve independent fluctuations of $A$ and $s_{iy}$. One can now write down the classical Lagrangian,
\begin{eqnarray*}
L &=& \frac{m}{2}\sum\limits_i \left(\dot x^2 + A \dot s_y^2 + A \dot s_z^2\right)
+ \frac{QL_x^2}{2}\dot A^2 \\
&& - U\left(A, \{s_{iy}, s_{iz}\}\right) - \overline{P_t}L_xA
\end{eqnarray*}
where we have introduced an artificial mass $Q$ for the new degree of freedom, setting the timescale of the area's fluctuations. Introducing the canonically conjugated momenta $p_{ix} \equiv m\dot x, \pi_{i\alpha} \equiv mA \dot s_{i\alpha}, \pi_A \equiv QL_x^2\dot A$, with $\alpha = \{y,z\}$, one can derive the Hamiltonian by means of a Legendre transformation
\begin{eqnarray*}
\mathcal{H} &=& \sum\limits_i\left(\frac{p_{ix}^2}{2m} + \frac{\pi_{iy}^2}{2mA} + \frac{\pi_{iz}^2}{2mA}\right) + \frac{\pi_A^2}{2QL_x^2} \\
&& + U\left(A, \{s_{iy}, s_{iz}\}\right) + \overline{P_t}L_xA. 
\end{eqnarray*}
The Hamilton equations of motion read
\begin{eqnarray*}
\dot x_i &=& \frac{p_{ix}}{m} \qquad \dot p_{ix} = F_{ix} \\
\dot s_{i\alpha} &=& \frac{\pi_{i\alpha}}{mA} \qquad \dot \pi_{i\alpha} = F_{i\alpha}\sqrt{A} \\
\dot A &=& \frac{\pi_A}{QL_x^2} \qquad \dot \pi_A = L_x\left(P_t - \overline{P_t}\right) 
\end{eqnarray*}
with $P_t \equiv (\sigma_{yy} + \sigma_{zz})/2$, where
\begin{equation*}
\sigma_{\alpha\alpha} = \frac{1}{AL_x}\sum\limits_i\left(\frac{\pi_{i\alpha}^2}{mA} + F_{i\alpha} s_{i\alpha} \sqrt{A}\right)
\end{equation*}
are the diagonal entries in the pressure tensor, that can be calculated with the virial theorem.

Following \citet{TUCKERMAN92}, we split the Liouville operator, $i\op L$, into
a sum of simpler operators
\begin{eqnarray*}
i\op L_1 &=& \sum\limits_i F_{ix} \frac{\partial}{\partial p_{ix}} + F_{iy} \sqrt{A} \frac{\partial}{\partial \pi_{iy}} +
F_{iz} \sqrt{A} \frac{\partial}{\partial \pi_{iz}} \nonumber \\
i\op L_2 &=& L_x \left (P_t - \overline{P_t}\right) \frac{\partial}{\partial \pi_A} \nonumber \\
i\op L_3 &=& \frac{\pi_A}{QL_x^2} \frac{\partial}{\partial A} \nonumber \\
i\op L_4 &=& \sum\limits_i \frac{p_{ix}}{m} \frac{\partial}{\partial x_i} + \frac{\pi_{iy}}{mA} \frac{\partial}{\partial s_{iy}} +
\frac{\pi_{iz}}{mA} \frac{\partial}{\partial s_{iz}}.
\end{eqnarray*}
We approximate the unitary time evolution operator by the Trotter factorization, yielding
\begin{eqnarray*}
U(\Delta t) &=& e^{i\op L\Delta t} = e^{i(\op L_1 + \op L_2 + \op L_3 + \op L_4)\Delta t} \\
&\approx& 
e^{i\op L_1 \frac{\Delta t}{2}}
e^{i\op L_2 \frac{\Delta t}{2}}
e^{i\op L_3 \frac{\Delta t}{2}}
e^{i\op L_4 \Delta t}
e^{i\op L_3 \frac{\Delta t}{2}}
e^{i\op L_2 \frac{\Delta t}{2}}
e^{i\op L_1 \frac{\Delta t}{2}}
\end{eqnarray*}
Applying these operators one after another from the right to the left onto the phase space vector, one obtains a symplectic integration algorithm in the canonically conjugated quantities:
\begin{enumerate}
\item 
$p_{ix}(\frac{\Delta t}{2}) = p_{ix}(0) + F_{ix} \frac{\Delta t}{2}$ \\
$\pi_{i\alpha}(\frac{\Delta t}{2}) = \pi_{i\alpha}(0) + F_{i\alpha}\sqrt{A(0)} \frac{\Delta t}{2}$
\item 
$\pi_A(\frac{\Delta t}{2}) = \pi_A(0) + L_x (P_t - \overline{P_t}) \frac{\Delta t}{2}$
\item
$A(\frac{\Delta t}{2}) = A(0) + \frac{\pi_A(\Delta t/2)}{QL_x^2} \frac{\Delta t}{2}$
\item
$x_i(\Delta t) = x_i(0) + \frac{p_{ix}(\Delta t/2)}{m}\Delta t$ \\
$s_{i\alpha}(\Delta t) = s_{i\alpha}(0) + \frac{\pi_{i\alpha}(\Delta t/2)}{mA}\Delta t$
\item
$A(\Delta t) = A(\frac{\Delta t}{2}) + \frac{\pi_A(\Delta t/2)}{QL_x^2} \frac{\Delta t}{2}$
\item
$\pi_A(\Delta t) = \pi_A(\frac{\Delta t}{2}) + L_x (P_t - \overline{P_t}) \frac{\Delta t}{2}$
\item
$p_{ix}(\Delta t) = p_{ix}(\frac{\Delta t}{2}) + F_{ix} \frac{\Delta t}{2}$ \\
$\pi_{i\alpha}(\Delta t) = \pi_{i\alpha}(\frac{\Delta t}{2}) + F_{i\alpha}\sqrt{A(\Delta t)} \frac{\Delta t}{2}$
\end{enumerate}
To simplify the usage of this algorithm the scaled coordinates, $s_{i\alpha}$, are finally substituted by the real coordinates, i.e.
\begin{eqnarray*}
y_i(t) = s_{iy}(t) \sqrt{A(t)} &\qquad&
p_{iy}(t) = \frac{\pi_{iy}(t)}{\sqrt{A(t)}} \\
z_i(t) = s_{iz}(t) \sqrt{A(t)} &\qquad&
p_{iz}(t) = \frac{\pi_{iz}(t)}{\sqrt{A(t)}}.
\end{eqnarray*}
This invokes some rescaling steps in the final algorithm. Hence:
\begin{enumerate}

\item Calculation of the temporary momenta using the old forces:
\begin{eqnarray*}
p_{ix}(\Delta t/2) &=& p_{ix}(0) + F_{ix} \frac{\Delta t}{2} \\
p_{i\alpha}' &\equiv& \frac{\pi_{i\alpha}(\Delta t/2)}{\sqrt{A(0)}} = p_{i\alpha}(0) + F_{i\alpha} \frac{\Delta t}{2}.
\end{eqnarray*}

\item Calculation of the intermediate area momentum using the tangential pressure $P_t$ evaluated
with the old forces $F(0)$ and the new temporary momenta $p_i'$:
\begin{equation*}
\pi_A(\Delta t/2) = \pi_A(0) + L_x (P_t - \overline{P_t}) \frac{\Delta t}{2}
\end{equation*}

\item First half of the integration of the area:
\begin{equation*}
A(\Delta t/2) = A(0) + \frac{\pi_A(\Delta t/2)}{QL_x^2} \frac{\Delta t}{2}
\end{equation*}

\item Integration of the particle coordinates:
\begin{eqnarray*}
x_i(\Delta t) &=& x_i(0) + \frac{p_{ix}(\Delta t/2)}{m}\Delta t \\ \nonumber
\alpha_i' &\equiv& s_{i\alpha}(\Delta t) \sqrt{A(0)} = \alpha_i(0) + \frac{A(0)}{A(\Delta t/2)} \frac{p_{i\alpha}'}{m} \Delta t
\end{eqnarray*}

\item Second half of the integration of the area:
\begin{equation*}
A(\Delta t) = A(\Delta t/2) + \frac{\pi_A(\Delta t/2)}{QL_x^2} \frac{\Delta t}{2}
\end{equation*}

\item Rescaling of the coordinates according to:
\begin{equation*}
\alpha_i(\Delta t) = \frac{\sqrt{A(t)}}{\sqrt{A(0)}} \alpha_i' \qquad p_{i\alpha}'' \equiv \frac{\sqrt{A(0)}}{\sqrt{A(\Delta t)}} p_{i\alpha}'
\end{equation*}

\item Recalculation of the new forces using the new coordinates and recalculation of the
pressure tensor using the new forces and the temporary momenta $p_{i\alpha}''$.

\item Calculation of the final area momentum using the tangential pressure $P_t$ evaluated
with the new forces $F(\Delta t)$ and the temporary momenta $p_{i\alpha}''$:
\begin{equation*}
\pi_A(\Delta t) = \pi_A(\Delta t/2) + L_x (P_t - \overline{P_t}) \frac{\Delta t}{2}
\end{equation*}

\item Final integration of the particle's momentum:
\begin{eqnarray*}
p_{ix}(\Delta t) &=& p_{ix}(\Delta t/2) + F_{ix} \frac{\Delta t}{2} \\ \nonumber
\pi_{i\alpha}(\Delta t) &=& p_{i\alpha}'' + F_{i\alpha} \frac{\Delta t}{2}
\end{eqnarray*}
\end{enumerate}

Up to now the integration algorithm has been formulated in the microcanonical ensemble, where the total energy is conserved. The switch to the $NP_tT$-ensemble is performed with the DPD thermostat for the particle interactions and with a Langevin thermostat for the area $A(t)$. The former is described in section \ref{model} and the latter is achieved via the replacement\cite{Kolb99}
\begin{eqnarray*}
L_x (P_t - \overline{P_t}) \frac{\Delta t}{2} &\to& L_x (P_t - \overline{P_t}) \frac{\Delta t}{2} \nonumber \\
&& - \gamma_A \frac{\pi_A}{QL_x^2} \frac{\Delta t}{2} + \sqrt{k_BT \gamma_A \Delta t / 2} \xi_A.
\end{eqnarray*}
In the last equation, $\xi_A$ is a random number drawn from a uniform distribution
with $\mv{\xi_A}=0$ and $\mv{\xi_A^2}=1$ and $\gamma_A$ is a friction coefficient.

In our simulations we have used the values $Q=0.0001, \gamma_A=0.1$ and $\overline{P_t}=0$, which corresponds to a simulation at vanishing lateral tension.

\section{Capillary waves and line tension}
\label{capillary-wave-appendix}
In this Appendix we briefly present the derivation of Eqs.~\eqref{line-spectrum} and \eqref{thd-line-tension}.  The statistical properties of the phase boundary follow from the capillary wave Hamiltonian
\begin{equation*}
\mathcal{H}_{\text{cap}} = \lambda \int\limits_0^L {\rm d}x \,
\sqrt{1+\left[h'(x)\right]^2}
\approx \lambda L + \frac{\lambda}{2} \int\limits_0^L {\rm d}x \,
\left[h'(x)\right]^2,
\label{CWH}
\end{equation*}
where $\lambda$ is the bare line tension, $L$ denotes the projected length of the line, and $h(x)$ represents the position of the phase boundary. The coordinate system is chosen such that the mean position of the boundary vanishes, i.e. $\mv{h(x)} = 0$. Expanding
\begin{equation*}
h(x) = \sum\limits_q h_q e^{iqx} \quad\text{with}\quad q = \frac{2\pi n}{L}
\end{equation*}
in a Fourier series with wave numbers $q$, $n \in \mathbb{Z}$, that are commensurate with the periodic boundary conditions, $\mathcal{H}_{\text{cap}}$ becomes diagonal in $q$-space and the different modes decouple. Hence, we rewrite $\mathcal{H}_{\text{cap}}$ in the form
\begin{equation*}
\mathcal{H}_{\text{cap}} = \lambda L + \mathcal{H}_f \quad \text{with}
\quad \mathcal{H}_f \equiv \frac{\lambda L}{2} \sum\limits_q \Abs{h_q}^2 q^2.
\end{equation*}
The Hamiltonian $\mathcal{H}_f$ is the starting point for all further calculations. On the one hand, the fluctuation power spectrum is readily obtained from this expression using the equipartition theorem\cite{Safran94}
\begin{equation*}
\mv{\Abs{h_q}^2} = \frac{k_BT}{\lambda L q^2}.
\end{equation*}
On the other hand, we calculate the free energy contribution, $\Delta F = F - \lambda L$, from the fluctuations of the boundary. The canonical partition function, $\mathcal{Z}$, involves a functional integral over all possible interface profiles, $h(x)$, which is equivalent to a functional integral over all complex Fourier coefficients, $h_q$:
\begin{eqnarray*}
\mathcal{Z} &\sim& \int \mathcal{D}\left[h_q\right]
\exp{\left(-\frac{\mathcal{H}_f}{k_BT}\right)} \\
&\sim& \prod\limits_q \int \frac{{\rm d} h_q}{L}
\exp{\left(-\frac{\lambda L}{2k_BT} \Abs{h_q}^2 q^2\right)}
\end{eqnarray*}
Since $h(x)$ is a real-valued function, the complex coefficients $h_q$ possess the Hermitian redundancy, i.e. $h_{-q} = \overline{h_q}$, so that we decompose the functional integral into separate integrals over the real and the imaginary parts of $h_q$ using only the positive modes, $q>0$. Thus,
\begin{equation*}
\prod\limits_q \int \frac{{\rm d} h_q}{L}  = \prod\limits_{q>0}
\int \frac{{\rm d}\left(\text{Re} \, h_q\right)}{L} \,
\frac{d\left(\text{Im} \, h_q\right)}{L}.
\end{equation*}
We evaluate $\mathcal{Z}$ by carrying out the Gaussian quadratures, yielding
\begin{equation*}
\mathcal{Z} \sim \prod\limits_{q>0} \frac{2\pi k_BT}{\lambda L}
\frac{1}{\left(qL\right)^2}
\end{equation*}
To compute the free energy $\Delta F=-k_BT\ln\mathcal{Z}$, the range of wave vectors has to be restricted. The simplest method is to introduce cut-offs for small $q$ as well as for high $q$. The small $q$-cutoff naturally arises from the periodic boundary conditions, i.e.,  $q_{\text{min}}=2\pi/L$. The UV-cutoff $q_{\text{max}} = 2\pi n_{\text{max}} / L = 2\pi / a, n_{\text{max}} = L/a$ is more difficult because it introduces an additional length scale. This length scale characterizes the smallest length scale, on which the fluctuations of the boundary line can be described by the capillary wave Hamiltonian. Using these two cut-offs, we obtain
\begin{eqnarray*}
\frac{\Delta F}{k_BT} &=& - \ln \mathcal{Z} = \sum\limits_{n=1}^{n_{\text{max}}}
\ln{\left( \frac{\lambda L}{2\pi k_BT} \cdot 4\pi^2 n^2\right)} \\
&=& n_{\text{max}} \ln{\left(\frac{2\pi\lambda L}{k_BT}\right)}
+ 2 \ln{\left(n_{\text{max}}!\right)}.
\end{eqnarray*}
$\Delta F$ is an extensive quantity, proportional to the size, $L$, of the system:
\begin{equation*}
\frac{\Delta F}{Lk_BT} = a^{-1}
\bigl[\ln{\left(\frac{2\pi\lambda L^3 }{ a^2k_BT }\right)} - 2 \bigr] \equiv \frac{C}{k_BT}
\end{equation*}
Hence, the total interfacial free energy is given by
\begin{equation*}
\label{line-free-energy}
F = \lambda L + \Delta F = (\lambda + C) L \equiv \sigma L.
\end{equation*}
Thus, the bare line tension, $\lambda$, in a one-dimensional system differs from the thermodynamic line tension, $\sigma$, by a quantity, $C$, which stems from the fluctuations of the contact line and which logarithmically depends on $L$ and $a$.


\begin{thebibliography}{104}
\bibitem{comment1} {An example of such a configuration is depicted in Fig.~\ref{fig:droplet-slab}b.}
\end{thebibliography}


\end{document}